\documentclass[aps,prl,reprint,twocolumn,longbibliography,superscriptaddress,floatfix]{revtex4-2}

\usepackage{amsmath}
\usepackage{txfonts}
\usepackage{graphicx}
\usepackage{bm}

\usepackage[dvipsnames]{xcolor}

\begin{document}

\title{ Visualization of Alternating Triangular Domains of Charge Density Waves in 2\textit{H}--NbSe$_2$ by Scanning Tunneling Microscopy }

\author{Shunsuke Yoshizawa}\email{YOSHIZAWA.Shunsuke@nims.go.jp}\affiliation{Center for Basic Research on Materials, National Institute for Materials Science, 1-2-1 Sengen, Tsukuba, Ibaraki 305-0047, Japan}
\author{Keisuke Sagisaka}\affiliation{Center for Basic Research on Materials, National Institute for Materials Science, 1-2-1 Sengen, Tsukuba, Ibaraki 305-0047, Japan}
\author{Hideaki Sakata}\affiliation{Department of Physics, Tokyo University of Science, 1-3 Kagurazaka, Shinjuku, Tokyo 162-8601, Japan}

\date{\today}

\begin{abstract}
The charge density wave (CDW) state of 2\textit{H}--NbSe$_2$ features commensurate domains separated by domain boundaries accompanied by phase slips known as discommensurations.
We have unambiguously visualized the structure of CDW domains using a displacement-field measurement algorithm on a scanning tunneling microscopy image.
Each CDW domain is delimited by three vertices and three edges of discommensurations and is designated by a triplet of integers whose sum identifies the types of commensurate structure.
The observed structure is consistent with the alternating triangular tiling pattern predicted by a phenomenological Landau theory.
The domain shape is affected by crystal defects and also by topological defects in the CDW phase factor.
Our results provide a foundation for a complete understanding of the CDW state and its relation to the superconducting state.
\end{abstract}

\maketitle


The charge density wave (CDW) is a cooperative phenomenon characteristic of low-dimensional metals \cite{Gruner1988,Zhu2015}. 
CDWs that coexist with other ordered states, exemplified by superconductivity, are of particular interest in view of the interplay between multiple orders \cite{Gabovich2001}. 
Notable examples are CDWs in cuprate superconductors \cite{Tranquada1995,Lee2022}, kagome superconductors \cite{Wang2021,Yu2021} and transition metal dichalcogenides (TMDCs) \cite{Manzeli2017}.
2\textit{H}--NbSe$_2$ is a prototypical TMDC that exhibits both the CDW and superconducting states.
It undergoes a triple-Q CDW transition, where three equivalent CDW wavevectors are $120^\circ$ apart, at a temperature of $T_\mathrm{CDW} \sim30$ K. 
Although neutron and X-ray diffraction experiments indicate an incommensurate periodicity \cite{Moncton1975,Du2000}, scanning tunneling microscopy (STM) studies have unveiled locally $3 \times 3$ commensurate domains \cite{Iwaya2003,Soumyanarayanan2013,Chatterjee2015,Guster2019,Gye2019,Oh2020}.
Domain boundaries serve as discommensurations, where the phase of the CDW shifts over a short distance \cite{McMillan1976}. 
Superconductivity occurs below the critical temperature of $T_\mathrm{c} \sim7$ K.
A recent STM study \cite{Liu2021} reveals the existence of the Cooper-pair density wave (PDW) state, which shares the same periodicity as the CDW with a constant phase difference.
These two modulated states have a common domain boundary. 
The results of another study \cite{Sanna2022} imply that the superconductivity competes with the CDW depending on the commensurate local structure. 
To better understand the interplay between the CDW and superconductivity, precise knowledge of the domain structure of the CDW is indispensable.

\begin{figure}
\includegraphics[width=85mm]{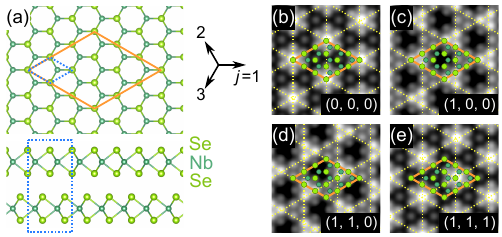}
\caption{\label{fig-model}
(a) Top view (upper panel) and front view (lower panel) of the crystal structure of 2\textit{H}--NbSe$_2$ \cite{Momma2011}.
The unit cell is depicted by a dotted rhombus (top view) and a dotted rectangle (front view).
The solid rhombus indicates the $3 \times 3$ unit cell of the commensurate CDW.
(b)--(e) Simulated CDW images with several combinations of indices $(n_1, n_2, n_3)$.
The dotted lines indicate the maxima of CDW components.
The solid rhombus indicates the $3 \times 3$ unit cell of the CDW and is fixed to the atomic lattice to show the relative position of the CDW maxima.
}\end{figure}

Despite extensive study, the domain structure of the CDW of 2\textit{H}--NbSe$_2$ remains elusive.
As shown in Fig. \ref{fig-model}(a), the crystal structure of 2\textit{H}--NbSe$_2$ consists of Nb atoms forming a triangular lattice and Se atoms protruding up and down from the Nb plane.
STM observations and density functional theory (DFT) calculations have identified two types of commensurate domain with distinct local structures \cite{Lian2018,Gye2019,Sanna2022}. 
One is characterized by the CDW maxima at Se sites, referred to as the chalcogen-centered (CC) structure.
The other is characterized by the CDW maxima at hollow sites and is referred to as the hollow-centered (HC) structure.
Gye \textit{et al}. have proposed a complex structure of these CDW domains \cite{Gye2019}, where HC domains are classified into nine types depending on the phase of modulation, forming isolated hexagonal patches, and CC domains surround HC domains and form a honeycomb network.
However, Gye \textit{et al}. did not provide a nonvisual method to identify the domain boundary.
Furthermore, they focused on the domain of only the HC structure and treated the CC structure as a type of discommensuration. 
This asymmetric treatment of the two commensurate structures and the ambiguity in domain identification may have prevented a simpler understanding of the domain structure.
Other groups reported the detection of discommensurations as steep changes in the phase factor of the CDW extracted from topographic images \cite{Pasztor2019,Liu2021}, but the nonvisual determination of the shape and the type of domains remained unresolved.

Here, we present a systematic approach to visualize the CDW domain structure of 2\textit{H}--NbSe$_2$ using topographic images obtained by STM.
Through a simple simulation, we demonstrate that the local CDW structure is characterized by a triplet of real numbers, denoted as $(n_1, n_2, n_3)$.
In each commensurate domain, the triplet takes integer values and their sum modulo 3 determines the type of commensurate structure.
We determined the spatial variation of $(n_1, n_2, n_3)$ by measuring the displacement fields of the CDW modulations relative to the atomic lattice in a topographic image, thereby revealing the domain structure without relying on visual interpretation.
Each CDW domain is bounded by three vertices and three edges, consistent with the alternating triangular tiling predicted by an empirical Landau theory.
The deformation of the domain shape from an equilateral triangle is attributed to the pinning of discommensurations by crystalline defects.

We used an ultrahigh-vacuum (UHV) cryogenic STM system based on the model USM-1300 from Unisoku Co. Ltd. 
The STM tip was fabricated by focused ion beam from a mechanically sharpened Pt--Ir wire and conditioned on a clean Au(111) surface prior to the experiment.
The single-crystal 2\textit{H}--NbSe$_2$ was grown by the chemical vapor transport method using iodine as the transport agent.  
The sample was cleaved in UHV at room temperature and immediately transferred to the STM head kept at a low temperature.
All the STM data presented in this paper were recorded in a constant-current mode at 4.5 K in zero magnetic field.
The resistivity measurement of a sample in the same batch gives the residual resistivity ratio of 42, $T_\mathrm{c} = 7.3$ K, and $T_\mathrm{CDW} \simeq 30$ K.

\begin{figure}
\includegraphics[width=85mm]{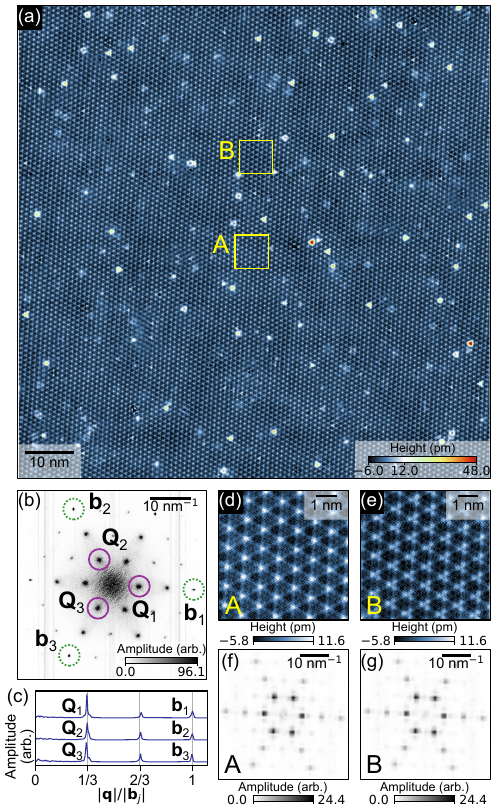}
\caption{\label{fig-stm}
(a) Topographic image of a 100 nm $\times$ 100 nm area of a cleaved surface of 2\textit{H}--NbSe$_2$.
The data were recorded at a resolution of 2048 $\times$ 2048 but were resampled to 1024 $\times$ 1024 to reduce the file size.
The image was flattened by subtracting a third-order polynomial fit from each line.
The feedback condition was 500 pA at 50 mV.
(b) Fourier transform of the topographic image in (a). 
The dotted and solid circles indicate the spots of the atomic lattice ($\mathbf{b}_1$, $\mathbf{b}_2$, and $\mathbf{b}_3$) and those of the CDW ($\mathbf{Q}_1$, $\mathbf{Q}_2$, and $\mathbf{Q}_3$), respectively.
(c) Line profiles of the Fourier transform image along $\mathbf{b}_1$, $\mathbf{b}_2$, and $\mathbf{b}_3$ directions.
(d) and (e) Cropped images of the CC and HC domains indicated by the boxes labeled A and B in (a).
(f) and (g) Fourier transform images of cropped data of (d) and (e).
}\end{figure}

Figure \ref{fig-stm}(a) shows a high-resolution topographic image of a 100 nm $\times$ 100 nm area that clearly resolves the CDW modulations as well as the triangular lattice of Se atoms.
The image also displays several types of atomic-scale defect.
The Fourier transform of this image [Fig. \ref{fig-stm}(b)] displays sharp peaks at the periodicity of the crystal lattice ($\mathbf{b}_1$, $\mathbf{b}_2$, and $\mathbf{b}_3$) and slightly broader peaks at the periodicity of the CDW ($\mathbf{Q}_1$, $\mathbf{Q}_2$, and $\mathbf{Q}_3$).
The broad nature of the latter peaks reflects the finite size of CDW domains.
All the other peaks are considered as harmonics of these periodicities.
The line profiles in Fig. \ref{fig-stm}(c) indicate that the length of $\mathbf{Q}_j$ is approximately one-third that of $\mathbf{b}_j$.
Upon closer examination of the topographic image, we observe a difference in local atomic-scale structure.
Figures \ref{fig-stm}(d) and \ref{fig-stm}(e) depict enlarged images cropped from the regions labeled A and B in Fig. \ref{fig-stm}(a), respectively.
The former displays a repeated star-shaped pattern that is characteristic of the CC structure, while the latter displays a repeated clover-shaped pattern of the HC structure \cite{Gye2019,Sanna2022}.
Figures \ref{fig-stm}(f) and \ref{fig-stm}(g) show the corresponding Fourier transforms.
The patterns are almost identical to each other at the pixel level.
This indicates that the CC and HC structures share the same periodic components but differ in their phase.

To investigate how the relative phase alters the topographic appearance, we present a simple simulation.
We approximate the charge density within a commensurate domain by the sum of two functions of position $\mathbf{r} = (x, y)$,
\begin{equation}
    \rho^{(n_1, n_2, n_3)}(\mathbf{r}) = \rho_{\mathrm{Se}}(\mathbf{r}) + c \rho_\mathrm{CDW}^{(n_1, n_2, n_3)}( \mathbf{r}).
\end{equation}
Here,
$ \rho_{\mathrm{Se}}(\mathbf{r}) = \operatorname{Re} \sum_{j=1}^3 \exp(i\mathbf{b}_j \cdot \mathbf{r}) $ 
represents the lattice modulations with maxima at Se sites, and
$ \rho_\mathrm{CDW}^{(n_1, n_2, n_3)}( \mathbf{r}) = \operatorname{Re} \sum_{j=1}^3 \exp[i (\mathbf{Q}_j \cdot \mathbf{r} - 2\pi n_j/3)] $ 
provides the CDW modulations with commensurate wavevectors $\mathbf{Q}_j = (1/3)\mathbf{b}_j$.
The indices $j = 1, 2, 3$ represent the directions.
The constant $c$ controls the amplitude of the CDW modulations relative to the lattice modulations, and we set it to be $\sim 2.5$.
The CDW modulations have phase offsets controlled by $n_j$.
Increasing $n_j$ by 1 translates the $j$-th component of the CDW by a lattice constant, and increasing it by 3 restores the CDW to its initial position.
The simulated topographic images $\rho^{(n_1, n_2, n_3)}(\mathbf{r})$ for several combinations of $(n_1, n_2, n_3)$ are shown in Figs. \ref{fig-model}(b)-\ref{fig-model}(e).
In the case of $(0, 0, 0)$ [Fig. \ref{fig-model}(b)], the CDW and the atomic lattice are in-phase, and CDW maxima are located at Se sites at the corner of the $3 \times 3$ cell.
This condition yields a topographic image of the CC structure.
In the case of $(1, 0, 0)$ [Fig. \ref{fig-model}(c)], CDW maxima are located at hollow sites, and the image shows the clover-shaped pattern of the HC structure.
In the case of $(1, 1, 0)$ [Fig. \ref{fig-model}(d)], CDW maxima are located at Nb sites. 
The pattern resembles that of the HC structure, but the orientation of the clover is the opposite.
In the case of $(1, 1, 1)$ [Fig. \ref{fig-model}(e)], the image again displays the CC structure, but CDW maxima are at Se sites different from those for $(0, 0, 0)$.
By examining the structure of various combinations of $(n_1, n_2, n_3)$ in this manner, we observe that the type of domain is determined by $ \lambda \equiv (\sum_{j=1}^3 n_j) \mod 3$, with $\lambda = 0$ corresponding to the CC structure and $\lambda = 1$ to the HC structure.
All the possible structures are presented in Section S1 of Supplemental Material (SM) \cite{Supplement}.

The precise determination of the triplet $(n_1, n_2, n_3)$ from the topographic image is essential for visualizing the domain structure. 
To achieve this, we utilize a displacement detection algorithm proposed by Lawler and Fujita {\it et al.} \cite{Lawler2010}.
 Assuming that the topographic image $z(\mathbf{r})$ contains a CDW modulation with periodicity $\mathbf{Q}_j$ that is altered by a slowly varying displacement field $\mathbf{u}_j(\mathbf{r})$, we obtain the displacement field using the relation:
\begin{equation}
    \exp[-i\mathbf{Q}_j\cdot\mathbf{u}_j(\mathbf{r})] \propto \sum_{\mathbf{r}'} z(\mathbf{r}') \exp(-i\mathbf{Q}_j\cdot\mathbf{r}') w(\mathbf{r} - \mathbf{r}'),
\end{equation}
where $w(\mathbf{r}) = (2\pi\sigma^2)^{-1} \exp[-|\mathbf{r}|^2 / (2\sigma^2)]$ is a two-dimensional Gaussian with spatial extent of $\sigma$.
We set $\sigma = 1$ nm in the present analysis.
From the phase component of the right-hand side, we obtain the (apparent) displacement field $u_j(\mathbf{r}) \equiv \mathbf{u}_j(\mathbf{r})\cdot\mathbf{Q}_j/|\mathbf{Q}_j|$ for each $j$.
However, this quantity includes an extrinsic image deformation, $\mathbf{v}(\mathbf{r})$, which originates from the creep effect of the piezoelectric scanner or the drift of the sample.
To eliminate this extrinsic effect, we extract $\mathbf{v}(\mathbf{r})$ by applying the Lawler--Fujita algorithm to the periodicities $\mathbf{b}_j$ of the crystal lattice.
We then obtain the intrinsic displacement fields of the CDW modulations by subtracting $v_j(\mathbf{r}) \equiv \mathbf{v}(\mathbf{r}) \cdot \mathbf{b}_j /|\mathbf{b}_j |$ from $u_j(\mathbf{r})$.
These intrinsic displacement fields are then divided by the interplanar spacing to obtain the spatial dependences of $(n_1, n_2, n_3)$.
A complete description of this procedure is presented in Section S2 of SM \cite{Supplement}.

\begin{figure}
\includegraphics[width=85mm]{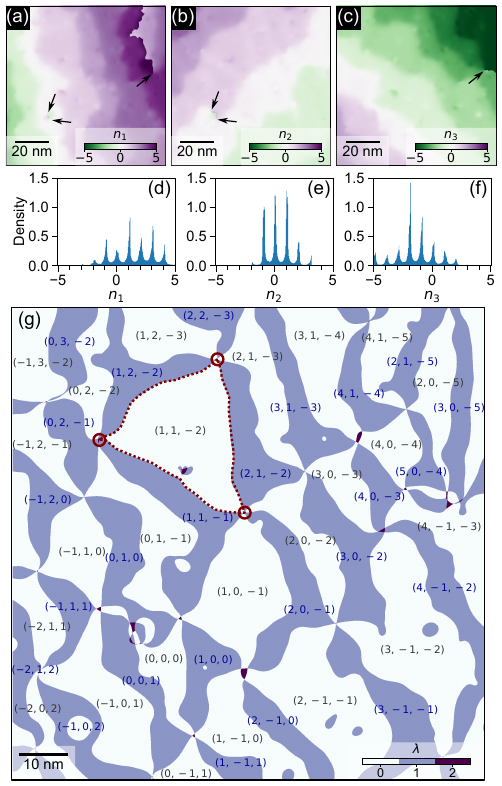}
\caption{\label{fig-domain}
(a)--(c) Images of $n_1(\mathbf{r})$, $n_2(\mathbf{r})$, and $n_3(\mathbf{r})$.
The arrows indicate the locations of topological defects.
(d)--(f) Histograms of $n_1(\mathbf{r})$, $n_2(\mathbf{r})$, and $n_3(\mathbf{r})$.
(g) Image of $\lambda(\mathbf{r})$.
The CC and HC domains are depicted as white and blue areas, respectively.
A triplet $(n_1, n_2, n_3)$ is written in each domain.
The circles and dotted curves are the vertices and edges of the $(0, 1, -1)$ CC domain, respectively.
}\end{figure}

Figures \ref{fig-domain}(a)--\ref{fig-domain}(c) show the images of $n_1(\mathbf{r})$, $n_2(\mathbf{r})$, and $n_3(\mathbf{r})$ determined from the topographic image in Fig. \ref{fig-stm}(a).
Each image displays a step-terrace structure, and the histograms in Figs. \ref{fig-domain}(d)--\ref{fig-domain}(f) show sharp peaks at integer values, highlighting the successful determination of $n_j$ values.
The results also show that the region is mostly covered by commensurate domains.
The domain structure was visualized in Fig. \ref{fig-domain}(g) by plotting
$\lambda(\mathbf{r}) \equiv \{\sum_j \operatorname{nint}[ n_j(\mathbf{r}) ]\} \mod 3,$
where $\operatorname{nint}$ denotes the nearest integer. 
The region is composed of alternating domains of $\lambda = 0$ (CC) and $\lambda = 1$ (HC), along with small areas with $\lambda = 2$ at the intersections of domain boundaries.
Each domain is enclosed by three vertices and three edges and is identified by a unique triplet of integers $(n_1, n_2, n_3)$.
The realization of this domain structure indicates that the discommensurations of the $\mathbf{Q}_1$, $\mathbf{Q}_2$, and $\mathbf{Q}_3$ components are not independent but are forced to intersect at a single vertex.
A direct comparison of the domain structure and the topographic image is shown in Fig. S7 of SM \cite{Supplement}.
We have confirmed the reproducibility of the present findings on another 2\textit{H}--NbSe$_2$ sample (see Section S4 of SM \cite{Supplement}).

We also evaluated the width of the discommensurations.
For each $j$, we defined the discommensurate region as the region where $n_j(\mathbf{r})$ is closer to a half-integer than to an integer.
We then estimated the width by dividing the area of the discommensurate region by the total length of the discommensuration.
Details are provided in Section S3 of SM \cite{Supplement}.
The estimated width was about 3 nm for all CDW components.
This is not an artefact but the real width of the discommensuration, because the value is larger than the broadening induced by the $\sigma$ of the Gaussian function $w(\mathbf{r})$.
This finding implies why Gye \textit{et al.} \cite{Gye2019} proposed a picture different from ours, where the HC domains were suggested to form isolated patches separated by continuous CC regions.
This interpretation arises when the discommensurate region in our definition is misassigned to the CC region.

\begin{figure}
\includegraphics[width=85mm]{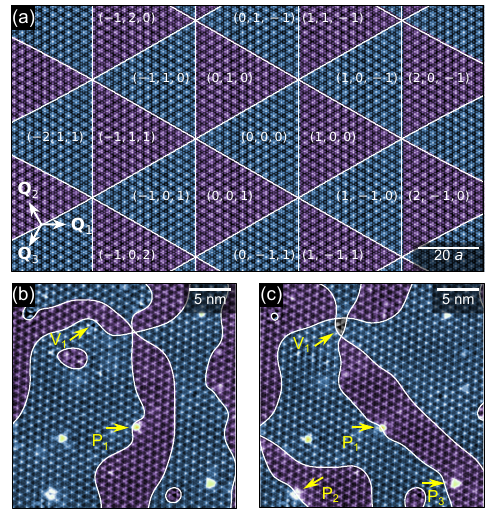}
\caption{\label{fig-pinning}
(a) Simulated topographic image of a $160a \times 90a$ region ($a$ is the lattice constant).
(b) Topographic image of a 25 nm $\times$ 25 nm region recorded after the initial cooling from room temperature to 4.5 K.
(c) Topographic image of the same region recorded after thermal cycles across $T_\mathrm{CDW}$.
The white curves represent the discommensurations.
The feedback condition was 500 pA at 50 mV.
The CC and HC domains are displayed in different colors for clarity.
}\end{figure}

The domain structure revealed here is consistent with the results of the studies based on the phenomenological Landau theory in the  1970s--1980s \cite{McMillan1976,Walker1981,Nakanishi1983,Nakanishi1986}.
In particular, McMillan's theory of discommensuration \cite{McMillan1976} was extended to describe the CDW states of TMDCs with a focus on 2\textit{H}--TaSe$_2$ \cite{Walker1981,Nakanishi1983,Nakanishi1986}.
These studies classified three possible commensurate structures, depending on the location of the threefold rotation axis, and showed that these structures were labeled by three integers, as we have found in our analysis.
Nakanishi and Shiba \cite{Nakanishi1983,Nakanishi1986} proposed a generic form of the free energy with many parameters and performed a numerical minimization for a parameter set that gives the situation where two types of commensurate structure become equally stable.
They predicted an alternating triangular tiling pattern of the two types of commensurate domain \cite{Nakanishi1986}.
Figure \ref{fig-pinning}(a) shows the CDW image simulated by minimizing Nakanishi-Shiba's free energy.
Details of the numerical simulation are shown in Section S5 of SM \cite{Supplement}.
Although the parameter choice was not optimized to describe 2\textit{H}--NbSe$_2$, the predicted pattern is essentially equivalent to the domain structure obtained in our experiment.
Note that the theoretical model was proposed before STM began to be used to observe CDWs \cite{Giambattista1988}, and it took nearly 40 years for its direct verification in actual material.

The observed domain structure is heavily deformed compared with the predicted equilateral triangles. 
This deformation is attributed to the pinning of discommensurations by defects in the crystal, as discussed in theoretical studies \cite{Nakanishi1979, Rice1981}. 
Figures \ref{fig-pinning}(b) and \ref{fig-pinning}(c) show the topography of the same region at 4.5 K before and after a thermal cycle across $T_\mathrm{CDW}$.
The thermal cycling has significantly altered the domain structure.
Before the thermal cycle, a protruding defect $\mathrm{P}_1$ was located on a discommensuration. 
After the thermal cycle, $\mathrm{P}_1$ and $\mathrm{P}_2$ were on discommensurations and $\mathrm{P}_3$ was also close to a discommensuration.
This observation indicates that these protruding defects potentially act as pinning centers.
It is reasonable that only some of the defects pin discommensurations, because the present sample surface has a large defect density compared with the spacing between discommensurations.
A void-like defect $\mathrm{V}_1$ also appears to pin discommensurations.
The exact roles of the different types of defect remain unclear in the present study. 
In addition, it is still uncertain why defects locally induce CDW above $T_\mathrm{CDW}$ \cite{Arguello2014,Oh2020}, but attract discommensurations below $T_\mathrm{CDW}$.
Although elucidating these points is beyond the scope of this study, our domain visualization method should be crucial for such investigations. 

The changes in the domain structure induced by thermal cycling suggest that the observed domain structures correspond to metastable states even at 4.5 K. 
Smooth relaxation to the ground state was probably prevented by the presence of topological defects, as the density of topological defects depends on the cooling rate \cite{Kibble2007}, a parameter not controlled in this study.
Topological defects in the CDW state of 2\textit{H}--NbSe$_2$ are known to be excited as vortex--antivortex pairs in the phase factor of the CDW \cite{Okamoto2015,Pasztor2019}. 
In our case, vortices and antivortices exist at the terminations of the discontinuities in the image of $n_1(\mathbf{r})$ [Fig. \ref{fig-domain}(a)]. 
Interestingly, we find that topological defects also exist in $n_2(\mathbf{r})$ and $n_3(\mathbf{r})$ at the same locations as those in $n_1(\mathbf{r})$ [Figs. \ref{fig-domain}(b) and \ref{fig-domain}(c)]. 
This highlights the fact that the topological defects in this system cannot exist independently in a single CDW component, but always appear as vortex--antivortex pairs in any two of the three components.
This can prevent $\sum_j n_j$ from changing by more than one around the vortex, thereby avoiding the formation of a region with $\lambda = 2$.
Owing to this intertwined nature of the three CDW components, vortex--antivortex pairs of each CDW component cannot annihilate independently but must do so simultaneously.
This restriction may hinder the relaxation to the ground-state domain structure free of topological defects.

In conclusion, our low-temperature STM observations, combined with systematic data analysis, have provided an accurate depiction of the domain structure of the CDW state in 2\textit{H}--NbSe$_2$. 
Our findings reveal that each CDW domain is characterized by a triplet of integers, and the domains form an alternating triangular tiling that is deformed by the presence of crystalline and topological defects.
Our work provides a basis for understanding the CDW transition in 2\textit{H}--NbSe$_2$ and other TMDC systems, and prompts further investigation of the interplay between the CDW and superconductivity.


\begin{acknowledgments}
The authors thank Y. Hattori and T. Terashima for carrying out transport measurements of 2\textit{H}--NbSe$_2$ samples.
This work was supported by JSPS KAKENHI (Grant Numbers 20H05277, 21K18898, and 21H01817).
\end{acknowledgments}


\begin{thebibliography}{35}%
\makeatletter
\providecommand \@ifxundefined [1]{%
 \@ifx{#1\undefined}
}%
\providecommand \@ifnum [1]{%
 \ifnum #1\expandafter \@firstoftwo
 \else \expandafter \@secondoftwo
 \fi
}%
\providecommand \@ifx [1]{%
 \ifx #1\expandafter \@firstoftwo
 \else \expandafter \@secondoftwo
 \fi
}%
\providecommand \natexlab [1]{#1}%
\providecommand \enquote  [1]{``#1''}%
\providecommand \bibnamefont  [1]{#1}%
\providecommand \bibfnamefont [1]{#1}%
\providecommand \citenamefont [1]{#1}%
\providecommand \href@noop [0]{\@secondoftwo}%
\providecommand \href [0]{\begingroup \@sanitize@url \@href}%
\providecommand \@href[1]{\@@startlink{#1}\@@href}%
\providecommand \@@href[1]{\endgroup#1\@@endlink}%
\providecommand \@sanitize@url [0]{\catcode `\\12\catcode `\$12\catcode
  `\&12\catcode `\#12\catcode `\^12\catcode `\_12\catcode `\%12\relax}%
\providecommand \@@startlink[1]{}%
\providecommand \@@endlink[0]{}%
\providecommand \url  [0]{\begingroup\@sanitize@url \@url }%
\providecommand \@url [1]{\endgroup\@href {#1}{\urlprefix }}%
\providecommand \urlprefix  [0]{URL }%
\providecommand \Eprint [0]{\href }%
\providecommand \doibase [0]{https://doi.org/}%
\providecommand \selectlanguage [0]{\@gobble}%
\providecommand \bibinfo  [0]{\@secondoftwo}%
\providecommand \bibfield  [0]{\@secondoftwo}%
\providecommand \translation [1]{[#1]}%
\providecommand \BibitemOpen [0]{}%
\providecommand \bibitemStop [0]{}%
\providecommand \bibitemNoStop [0]{.\EOS\space}%
\providecommand \EOS [0]{\spacefactor3000\relax}%
\providecommand \BibitemShut  [1]{\csname bibitem#1\endcsname}%
\let\auto@bib@innerbib\@empty
\bibitem [{\citenamefont {Gr\"uner}(1988)}]{Gruner1988}%
  \BibitemOpen
  \bibfield  {author} {\bibinfo {author} {\bibfnamefont {G.}~\bibnamefont
  {Gr\"uner}},\ }\bibfield  {title} {\bibinfo {title} {The dynamics of
  charge-density waves},\ }\href {https://doi.org/10.1103/revmodphys.60.1129}
  {\bibfield  {journal} {\bibinfo  {journal} {Rev. Mod. Phys.}\ }\textbf
  {\bibinfo {volume} {60}},\ \bibinfo {pages} {1129} (\bibinfo {year}
  {1988})}\BibitemShut {NoStop}%
\bibitem [{\citenamefont {Zhu}\ \emph {et~al.}(2015)\citenamefont {Zhu},
  \citenamefont {Cao}, \citenamefont {Zhang}, \citenamefont {Plummer},\ and\
  \citenamefont {Guo}}]{Zhu2015}%
  \BibitemOpen
  \bibfield  {author} {\bibinfo {author} {\bibfnamefont {X.}~\bibnamefont
  {Zhu}}, \bibinfo {author} {\bibfnamefont {Y.}~\bibnamefont {Cao}}, \bibinfo
  {author} {\bibfnamefont {J.}~\bibnamefont {Zhang}}, \bibinfo {author}
  {\bibfnamefont {E.~W.}\ \bibnamefont {Plummer}},\ and\ \bibinfo {author}
  {\bibfnamefont {J.}~\bibnamefont {Guo}},\ }\bibfield  {title} {\bibinfo
  {title} {Classification of charge density waves based on their nature},\
  }\href {https://doi.org/10.1073/pnas.1424791112} {\bibfield  {journal}
  {\bibinfo  {journal} {Proc. Natl. Acad. Sci. U.S.A.}\ }\textbf {\bibinfo
  {volume} {112}},\ \bibinfo {pages} {2367} (\bibinfo {year}
  {2015})}\BibitemShut {NoStop}%
\bibitem [{\citenamefont {Gabovich}\ \emph {et~al.}(2001)\citenamefont
  {Gabovich}, \citenamefont {Voitenko}, \citenamefont {Annett},\ and\
  \citenamefont {Ausloos}}]{Gabovich2001}%
  \BibitemOpen
  \bibfield  {author} {\bibinfo {author} {\bibfnamefont {A.~M.}\ \bibnamefont
  {Gabovich}}, \bibinfo {author} {\bibfnamefont {A.~I.}\ \bibnamefont
  {Voitenko}}, \bibinfo {author} {\bibfnamefont {J.~F.}\ \bibnamefont
  {Annett}},\ and\ \bibinfo {author} {\bibfnamefont {M.}~\bibnamefont
  {Ausloos}},\ }\bibfield  {title} {\bibinfo {title} {Charge- and
  spin-density-wave superconductors},\ }\href
  {https://doi.org/10.1088/0953-2048/14/4/201} {\bibfield  {journal} {\bibinfo
  {journal} {Supercond. Sci. Technol.}\ }\textbf {\bibinfo {volume} {14}},\
  \bibinfo {pages} {R1} (\bibinfo {year} {2001})}\BibitemShut {NoStop}%
\bibitem [{\citenamefont {Tranquada}\ \emph {et~al.}(1995)\citenamefont
  {Tranquada}, \citenamefont {Sternlieb}, \citenamefont {Axe}, \citenamefont
  {Nakamura},\ and\ \citenamefont {Uchida}}]{Tranquada1995}%
  \BibitemOpen
  \bibfield  {author} {\bibinfo {author} {\bibfnamefont {J.~M.}\ \bibnamefont
  {Tranquada}}, \bibinfo {author} {\bibfnamefont {B.~J.}\ \bibnamefont
  {Sternlieb}}, \bibinfo {author} {\bibfnamefont {J.~D.}\ \bibnamefont {Axe}},
  \bibinfo {author} {\bibfnamefont {Y.}~\bibnamefont {Nakamura}},\ and\
  \bibinfo {author} {\bibfnamefont {S.}~\bibnamefont {Uchida}},\ }\bibfield
  {title} {\bibinfo {title} {Evidence for stripe correlations of spins and
  holes in copper oxide superconductors},\ }\href
  {https://doi.org/10.1038/375561a0} {\bibfield  {journal} {\bibinfo  {journal}
  {Nature (London)}\ }\textbf {\bibinfo {volume} {375}},\ \bibinfo {pages}
  {561} (\bibinfo {year} {1995})}\BibitemShut {NoStop}%
\bibitem [{\citenamefont {Lee}\ \emph {et~al.}(2022)\citenamefont {Lee},
  \citenamefont {Huang}, \citenamefont {Johnson}, \citenamefont {Guo},
  \citenamefont {Husain}, \citenamefont {Mitrano}, \citenamefont {Lu},
  \citenamefont {Zakrzewski}, \citenamefont {de~la Pe{\~{n}}a}, \citenamefont
  {Peng}, \citenamefont {Huang}, \citenamefont {Lee}, \citenamefont {Jang},
  \citenamefont {Lee}, \citenamefont {Joe}, \citenamefont {Doriese},
  \citenamefont {Szypryt}, \citenamefont {Swetz}, \citenamefont {Chi},
  \citenamefont {Aczel}, \citenamefont {MacDougall}, \citenamefont {Kivelson},
  \citenamefont {Fradkin},\ and\ \citenamefont {Abbamonte}}]{Lee2022}%
  \BibitemOpen
  \bibfield  {author} {\bibinfo {author} {\bibfnamefont {S.}~\bibnamefont
  {Lee}}, \bibinfo {author} {\bibfnamefont {E.~W.}\ \bibnamefont {Huang}},
  \bibinfo {author} {\bibfnamefont {T.~A.}\ \bibnamefont {Johnson}}, \bibinfo
  {author} {\bibfnamefont {X.}~\bibnamefont {Guo}}, \bibinfo {author}
  {\bibfnamefont {A.~A.}\ \bibnamefont {Husain}}, \bibinfo {author}
  {\bibfnamefont {M.}~\bibnamefont {Mitrano}}, \bibinfo {author} {\bibfnamefont
  {K.}~\bibnamefont {Lu}}, \bibinfo {author} {\bibfnamefont {A.~V.}\
  \bibnamefont {Zakrzewski}}, \bibinfo {author} {\bibfnamefont {G.~A.}\
  \bibnamefont {de~la Pe{\~{n}}a}}, \bibinfo {author} {\bibfnamefont
  {Y.}~\bibnamefont {Peng}}, \bibinfo {author} {\bibfnamefont {H.}~\bibnamefont
  {Huang}}, \bibinfo {author} {\bibfnamefont {S.-J.}\ \bibnamefont {Lee}},
  \bibinfo {author} {\bibfnamefont {H.}~\bibnamefont {Jang}}, \bibinfo {author}
  {\bibfnamefont {J.-S.}\ \bibnamefont {Lee}}, \bibinfo {author} {\bibfnamefont
  {Y.~I.}\ \bibnamefont {Joe}}, \bibinfo {author} {\bibfnamefont {W.~B.}\
  \bibnamefont {Doriese}}, \bibinfo {author} {\bibfnamefont {P.}~\bibnamefont
  {Szypryt}}, \bibinfo {author} {\bibfnamefont {D.~S.}\ \bibnamefont {Swetz}},
  \bibinfo {author} {\bibfnamefont {S.}~\bibnamefont {Chi}}, \bibinfo {author}
  {\bibfnamefont {A.~A.}\ \bibnamefont {Aczel}}, \bibinfo {author}
  {\bibfnamefont {G.~J.}\ \bibnamefont {MacDougall}}, \bibinfo {author}
  {\bibfnamefont {S.~A.}\ \bibnamefont {Kivelson}}, \bibinfo {author}
  {\bibfnamefont {E.}~\bibnamefont {Fradkin}},\ and\ \bibinfo {author}
  {\bibfnamefont {P.}~\bibnamefont {Abbamonte}},\ }\bibfield  {title} {\bibinfo
  {title} {Generic character of charge and spin density waves in
  superconducting cuprates},\ }\href {https://doi.org/10.1073/pnas.2119429119}
  {\bibfield  {journal} {\bibinfo  {journal} {Proc. Natl. Acad. Sci. U.S.A.}\
  }\textbf {\bibinfo {volume} {119}},\ \bibinfo {pages} {e2119429119} (\bibinfo
  {year} {2022})}\BibitemShut {NoStop}%
\bibitem [{\citenamefont {Wang}\ \emph {et~al.}(2021)\citenamefont {Wang},
  \citenamefont {Chen}, \citenamefont {Yin}, \citenamefont {Ma}, \citenamefont
  {Pan}, \citenamefont {Yang}, \citenamefont {Ji}, \citenamefont {Wu},
  \citenamefont {Shan}, \citenamefont {Xu}, \citenamefont {Tu}, \citenamefont
  {Gong}, \citenamefont {Liu}, \citenamefont {Li}, \citenamefont {Uwatoko},
  \citenamefont {Dong}, \citenamefont {Lei}, \citenamefont {Sun},\ and\
  \citenamefont {Cheng}}]{Wang2021}%
  \BibitemOpen
  \bibfield  {author} {\bibinfo {author} {\bibfnamefont {N.~N.}\ \bibnamefont
  {Wang}}, \bibinfo {author} {\bibfnamefont {K.~Y.}\ \bibnamefont {Chen}},
  \bibinfo {author} {\bibfnamefont {Q.~W.}\ \bibnamefont {Yin}}, \bibinfo
  {author} {\bibfnamefont {Y.~N.~N.}\ \bibnamefont {Ma}}, \bibinfo {author}
  {\bibfnamefont {B.~Y.}\ \bibnamefont {Pan}}, \bibinfo {author} {\bibfnamefont
  {X.}~\bibnamefont {Yang}}, \bibinfo {author} {\bibfnamefont {X.~Y.}\
  \bibnamefont {Ji}}, \bibinfo {author} {\bibfnamefont {S.~L.}\ \bibnamefont
  {Wu}}, \bibinfo {author} {\bibfnamefont {P.~F.}\ \bibnamefont {Shan}},
  \bibinfo {author} {\bibfnamefont {S.~X.}\ \bibnamefont {Xu}}, \bibinfo
  {author} {\bibfnamefont {Z.~J.}\ \bibnamefont {Tu}}, \bibinfo {author}
  {\bibfnamefont {C.~S.}\ \bibnamefont {Gong}}, \bibinfo {author}
  {\bibfnamefont {G.~T.}\ \bibnamefont {Liu}}, \bibinfo {author} {\bibfnamefont
  {G.}~\bibnamefont {Li}}, \bibinfo {author} {\bibfnamefont {Y.}~\bibnamefont
  {Uwatoko}}, \bibinfo {author} {\bibfnamefont {X.~L.}\ \bibnamefont {Dong}},
  \bibinfo {author} {\bibfnamefont {H.~C.}\ \bibnamefont {Lei}}, \bibinfo
  {author} {\bibfnamefont {J.~P.}\ \bibnamefont {Sun}},\ and\ \bibinfo {author}
  {\bibfnamefont {J.-G.}\ \bibnamefont {Cheng}},\ }\bibfield  {title} {\bibinfo
  {title} {Competition between charge-density-wave and superconductivity in the
  kagome metal {RbV$_3$Sb$_5$}},\ }\href
  {https://doi.org/10.1103/physrevresearch.3.043018} {\bibfield  {journal}
  {\bibinfo  {journal} {Phys. Rev. Res.}\ }\textbf {\bibinfo {volume} {3}},\
  \bibinfo {pages} {043018} (\bibinfo {year} {2021})}\BibitemShut {NoStop}%
\bibitem [{\citenamefont {Yu}\ \emph {et~al.}(2021)\citenamefont {Yu},
  \citenamefont {Ma}, \citenamefont {Zhuo}, \citenamefont {Liu}, \citenamefont
  {Wen}, \citenamefont {Lei}, \citenamefont {Ying},\ and\ \citenamefont
  {Chen}}]{Yu2021}%
  \BibitemOpen
  \bibfield  {author} {\bibinfo {author} {\bibfnamefont {F.~H.}\ \bibnamefont
  {Yu}}, \bibinfo {author} {\bibfnamefont {D.~H.}\ \bibnamefont {Ma}}, \bibinfo
  {author} {\bibfnamefont {W.~Z.}\ \bibnamefont {Zhuo}}, \bibinfo {author}
  {\bibfnamefont {S.~Q.}\ \bibnamefont {Liu}}, \bibinfo {author} {\bibfnamefont
  {X.~K.}\ \bibnamefont {Wen}}, \bibinfo {author} {\bibfnamefont
  {B.}~\bibnamefont {Lei}}, \bibinfo {author} {\bibfnamefont {J.~J.}\
  \bibnamefont {Ying}},\ and\ \bibinfo {author} {\bibfnamefont {X.~H.}\
  \bibnamefont {Chen}},\ }\bibfield  {title} {\bibinfo {title} {Unusual
  competition of superconductivity and charge-density-wave state in a
  compressed topological kagome metal},\ }\href
  {https://doi.org/10.1038/s41467-021-23928-w} {\bibfield  {journal} {\bibinfo
  {journal} {Nat. Commun.}\ }\textbf {\bibinfo {volume} {12}},\ \bibinfo
  {pages} {3645} (\bibinfo {year} {2021})}\BibitemShut {NoStop}%
\bibitem [{\citenamefont {Manzeli}\ \emph {et~al.}(2017)\citenamefont
  {Manzeli}, \citenamefont {Ovchinnikov}, \citenamefont {Pasquier},
  \citenamefont {Yazyev},\ and\ \citenamefont {Kis}}]{Manzeli2017}%
  \BibitemOpen
  \bibfield  {author} {\bibinfo {author} {\bibfnamefont {S.}~\bibnamefont
  {Manzeli}}, \bibinfo {author} {\bibfnamefont {D.}~\bibnamefont
  {Ovchinnikov}}, \bibinfo {author} {\bibfnamefont {D.}~\bibnamefont
  {Pasquier}}, \bibinfo {author} {\bibfnamefont {O.~V.}\ \bibnamefont
  {Yazyev}},\ and\ \bibinfo {author} {\bibfnamefont {A.}~\bibnamefont {Kis}},\
  }\bibfield  {title} {\bibinfo {title} {{2D} transition metal
  dichalcogenides},\ }\href {https://doi.org/10.1038/natrevmats.2017.33}
  {\bibfield  {journal} {\bibinfo  {journal} {Nat. Rev. Mater.}\ }\textbf
  {\bibinfo {volume} {2}},\ \bibinfo {pages} {17033} (\bibinfo {year}
  {2017})}\BibitemShut {NoStop}%
\bibitem [{\citenamefont {Moncton}\ \emph {et~al.}(1975)\citenamefont
  {Moncton}, \citenamefont {Axe},\ and\ \citenamefont {DiSalvo}}]{Moncton1975}%
  \BibitemOpen
  \bibfield  {author} {\bibinfo {author} {\bibfnamefont {D.~E.}\ \bibnamefont
  {Moncton}}, \bibinfo {author} {\bibfnamefont {J.~D.}\ \bibnamefont {Axe}},\
  and\ \bibinfo {author} {\bibfnamefont {F.~J.}\ \bibnamefont {DiSalvo}},\
  }\bibfield  {title} {\bibinfo {title} {Study of superlattice formation in
  {2\textit{H}}--{NbSe}$_2$ and {2\textit{H}}--{TaSe}$_2$ by neutron
  scattering},\ }\href {https://doi.org/10.1103/physrevlett.34.734} {\bibfield
  {journal} {\bibinfo  {journal} {Phys. Rev. Lett.}\ }\textbf {\bibinfo
  {volume} {34}},\ \bibinfo {pages} {734} (\bibinfo {year} {1975})}\BibitemShut
  {NoStop}%
\bibitem [{\citenamefont {Du}\ \emph {et~al.}(2000)\citenamefont {Du},
  \citenamefont {Lin}, \citenamefont {Su}, \citenamefont {Tanner},
  \citenamefont {Hatton}, \citenamefont {Casa}, \citenamefont {Keimer},
  \citenamefont {Hill}, \citenamefont {Oglesby},\ and\ \citenamefont
  {Hohl}}]{Du2000}%
  \BibitemOpen
  \bibfield  {author} {\bibinfo {author} {\bibfnamefont {C.-H.}\ \bibnamefont
  {Du}}, \bibinfo {author} {\bibfnamefont {W.~J.}\ \bibnamefont {Lin}},
  \bibinfo {author} {\bibfnamefont {Y.}~\bibnamefont {Su}}, \bibinfo {author}
  {\bibfnamefont {B.~K.}\ \bibnamefont {Tanner}}, \bibinfo {author}
  {\bibfnamefont {P.~D.}\ \bibnamefont {Hatton}}, \bibinfo {author}
  {\bibfnamefont {D.}~\bibnamefont {Casa}}, \bibinfo {author} {\bibfnamefont
  {B.}~\bibnamefont {Keimer}}, \bibinfo {author} {\bibfnamefont {J.~P.}\
  \bibnamefont {Hill}}, \bibinfo {author} {\bibfnamefont {C.~S.}\ \bibnamefont
  {Oglesby}},\ and\ \bibinfo {author} {\bibfnamefont {H.}~\bibnamefont
  {Hohl}},\ }\bibfield  {title} {\bibinfo {title} {X-ray scattering studies of
  {2\textit{H}}--{NbSe}$_2$, a superconductor and charge density wave material,
  under high external magnetic fields},\ }\href
  {https://doi.org/10.1088/0953-8984/12/25/302} {\bibfield  {journal} {\bibinfo
   {journal} {J. Phys. Condens. Matter.}\ }\textbf {\bibinfo {volume} {12}},\
  \bibinfo {pages} {5361} (\bibinfo {year} {2000})}\BibitemShut {NoStop}%
\bibitem [{\citenamefont {Iwaya}\ \emph {et~al.}(2003)\citenamefont {Iwaya},
  \citenamefont {Hanaguri}, \citenamefont {Koizumi}, \citenamefont {Takaki},
  \citenamefont {Maeda},\ and\ \citenamefont {Kitazawa}}]{Iwaya2003}%
  \BibitemOpen
  \bibfield  {author} {\bibinfo {author} {\bibfnamefont {K.}~\bibnamefont
  {Iwaya}}, \bibinfo {author} {\bibfnamefont {T.}~\bibnamefont {Hanaguri}},
  \bibinfo {author} {\bibfnamefont {A.}~\bibnamefont {Koizumi}}, \bibinfo
  {author} {\bibfnamefont {K.}~\bibnamefont {Takaki}}, \bibinfo {author}
  {\bibfnamefont {A.}~\bibnamefont {Maeda}},\ and\ \bibinfo {author}
  {\bibfnamefont {K.}~\bibnamefont {Kitazawa}},\ }\bibfield  {title} {\bibinfo
  {title} {Electronic state of {NbSe}$_2$ investigated by {STM}/{STS}},\ }\href
  {https://doi.org/10.1016/s0921-4526(02)02309-8} {\bibfield  {journal}
  {\bibinfo  {journal} {Phys. B (Amsterdam, Neth.)}\ }\textbf {\bibinfo
  {volume} {329--333}},\ \bibinfo {pages} {1598} (\bibinfo {year}
  {2003})}\BibitemShut {NoStop}%
\bibitem [{\citenamefont {Soumyanarayanan}\ \emph {et~al.}(2013)\citenamefont
  {Soumyanarayanan}, \citenamefont {Yee}, \citenamefont {He}, \citenamefont
  {van Wezel}, \citenamefont {Rahn}, \citenamefont {Rossnagel}, \citenamefont
  {Hudson}, \citenamefont {Norman},\ and\ \citenamefont
  {Hoffman}}]{Soumyanarayanan2013}%
  \BibitemOpen
  \bibfield  {author} {\bibinfo {author} {\bibfnamefont {A.}~\bibnamefont
  {Soumyanarayanan}}, \bibinfo {author} {\bibfnamefont {M.~M.}\ \bibnamefont
  {Yee}}, \bibinfo {author} {\bibfnamefont {Y.}~\bibnamefont {He}}, \bibinfo
  {author} {\bibfnamefont {J.}~\bibnamefont {van Wezel}}, \bibinfo {author}
  {\bibfnamefont {D.~J.}\ \bibnamefont {Rahn}}, \bibinfo {author}
  {\bibfnamefont {K.}~\bibnamefont {Rossnagel}}, \bibinfo {author}
  {\bibfnamefont {E.~W.}\ \bibnamefont {Hudson}}, \bibinfo {author}
  {\bibfnamefont {M.~R.}\ \bibnamefont {Norman}},\ and\ \bibinfo {author}
  {\bibfnamefont {J.~E.}\ \bibnamefont {Hoffman}},\ }\bibfield  {title}
  {\bibinfo {title} {Quantum phase transition from triangular to stripe charge
  order in {NbSe}$_2$},\ }\href {https://doi.org/10.1073/pnas.1211387110}
  {\bibfield  {journal} {\bibinfo  {journal} {Proc. Natl. Acad. Sci. U.S.A.}\
  }\textbf {\bibinfo {volume} {110}},\ \bibinfo {pages} {1623} (\bibinfo {year}
  {2013})}\BibitemShut {NoStop}%
\bibitem [{\citenamefont {Chatterjee}\ \emph {et~al.}(2015)\citenamefont
  {Chatterjee}, \citenamefont {Zhao}, \citenamefont {Iavarone}, \citenamefont
  {Capua}, \citenamefont {Castellan}, \citenamefont {Karapetrov}, \citenamefont
  {Malliakas}, \citenamefont {Kanatzidis}, \citenamefont {Claus}, \citenamefont
  {Ruff}, \citenamefont {Weber}, \citenamefont {van Wezel}, \citenamefont
  {Campuzano}, \citenamefont {Osborn}, \citenamefont {Randeria}, \citenamefont
  {Trivedi}, \citenamefont {Norman},\ and\ \citenamefont
  {Rosenkranz}}]{Chatterjee2015}%
  \BibitemOpen
  \bibfield  {author} {\bibinfo {author} {\bibfnamefont {U.}~\bibnamefont
  {Chatterjee}}, \bibinfo {author} {\bibfnamefont {J.}~\bibnamefont {Zhao}},
  \bibinfo {author} {\bibfnamefont {M.}~\bibnamefont {Iavarone}}, \bibinfo
  {author} {\bibfnamefont {R.~D.}\ \bibnamefont {Capua}}, \bibinfo {author}
  {\bibfnamefont {J.~P.}\ \bibnamefont {Castellan}}, \bibinfo {author}
  {\bibfnamefont {G.}~\bibnamefont {Karapetrov}}, \bibinfo {author}
  {\bibfnamefont {C.~D.}\ \bibnamefont {Malliakas}}, \bibinfo {author}
  {\bibfnamefont {M.~G.}\ \bibnamefont {Kanatzidis}}, \bibinfo {author}
  {\bibfnamefont {H.}~\bibnamefont {Claus}}, \bibinfo {author} {\bibfnamefont
  {J.~P.~C.}\ \bibnamefont {Ruff}}, \bibinfo {author} {\bibfnamefont
  {F.}~\bibnamefont {Weber}}, \bibinfo {author} {\bibfnamefont
  {J.}~\bibnamefont {van Wezel}}, \bibinfo {author} {\bibfnamefont {J.~C.}\
  \bibnamefont {Campuzano}}, \bibinfo {author} {\bibfnamefont {R.}~\bibnamefont
  {Osborn}}, \bibinfo {author} {\bibfnamefont {M.}~\bibnamefont {Randeria}},
  \bibinfo {author} {\bibfnamefont {N.}~\bibnamefont {Trivedi}}, \bibinfo
  {author} {\bibfnamefont {M.~R.}\ \bibnamefont {Norman}},\ and\ \bibinfo
  {author} {\bibfnamefont {S.}~\bibnamefont {Rosenkranz}},\ }\bibfield  {title}
  {\bibinfo {title} {Emergence of coherence in the charge-density wave state of
  {2\textit{H}}-{NbSe}$_2$},\ }\href {https://doi.org/10.1038/ncomms7313}
  {\bibfield  {journal} {\bibinfo  {journal} {Nat. Commun.}\ }\textbf {\bibinfo
  {volume} {6}},\ \bibinfo {pages} {6313} (\bibinfo {year} {2015})}\BibitemShut
  {NoStop}%
\bibitem [{\citenamefont {Guster}\ \emph {et~al.}(2019)\citenamefont {Guster},
  \citenamefont {Rubio-Verd{\'{u}}}, \citenamefont {Robles}, \citenamefont
  {Zald{\'{\i}}var}, \citenamefont {Dreher}, \citenamefont {Pruneda},
  \citenamefont {Silva-Guill{\'{e}}n}, \citenamefont {Choi}, \citenamefont
  {Pascual}, \citenamefont {Ugeda}, \citenamefont {Ordej{\'{o}}n},\ and\
  \citenamefont {Canadell}}]{Guster2019}%
  \BibitemOpen
  \bibfield  {author} {\bibinfo {author} {\bibfnamefont {B.}~\bibnamefont
  {Guster}}, \bibinfo {author} {\bibfnamefont {C.}~\bibnamefont
  {Rubio-Verd{\'{u}}}}, \bibinfo {author} {\bibfnamefont {R.}~\bibnamefont
  {Robles}}, \bibinfo {author} {\bibfnamefont {J.}~\bibnamefont
  {Zald{\'{\i}}var}}, \bibinfo {author} {\bibfnamefont {P.}~\bibnamefont
  {Dreher}}, \bibinfo {author} {\bibfnamefont {M.}~\bibnamefont {Pruneda}},
  \bibinfo {author} {\bibfnamefont {J.~{\'{A}}.}\ \bibnamefont
  {Silva-Guill{\'{e}}n}}, \bibinfo {author} {\bibfnamefont {D.-J.}\
  \bibnamefont {Choi}}, \bibinfo {author} {\bibfnamefont {J.~I.}\ \bibnamefont
  {Pascual}}, \bibinfo {author} {\bibfnamefont {M.~M.}\ \bibnamefont {Ugeda}},
  \bibinfo {author} {\bibfnamefont {P.}~\bibnamefont {Ordej{\'{o}}n}},\ and\
  \bibinfo {author} {\bibfnamefont {E.}~\bibnamefont {Canadell}},\ }\bibfield
  {title} {\bibinfo {title} {Coexistence of elastic modulations in the charge
  density wave state of {2\textit{H}}--{NbSe}$_2$},\ }\href
  {https://doi.org/10.1021/acs.nanolett.9b00268} {\bibfield  {journal}
  {\bibinfo  {journal} {Nano Letters}\ }\textbf {\bibinfo {volume} {19}},\
  \bibinfo {pages} {3027} (\bibinfo {year} {2019})}\BibitemShut {NoStop}%
\bibitem [{\citenamefont {Gye}\ \emph {et~al.}(2019)\citenamefont {Gye},
  \citenamefont {Oh},\ and\ \citenamefont {Yeom}}]{Gye2019}%
  \BibitemOpen
  \bibfield  {author} {\bibinfo {author} {\bibfnamefont {G.}~\bibnamefont
  {Gye}}, \bibinfo {author} {\bibfnamefont {E.}~\bibnamefont {Oh}},\ and\
  \bibinfo {author} {\bibfnamefont {H.~W.}\ \bibnamefont {Yeom}},\ }\bibfield
  {title} {\bibinfo {title} {Topological landscape of competing charge density
  waves in {2\textit{H}}--{NbSe}$_2$},\ }\href
  {https://doi.org/10.1103/physrevlett.122.016403} {\bibfield  {journal}
  {\bibinfo  {journal} {Phys. Rev. Lett.}\ }\textbf {\bibinfo {volume} {122}},\
  \bibinfo {pages} {016403} (\bibinfo {year} {2019})}\BibitemShut {NoStop}%
\bibitem [{\citenamefont {Oh}\ \emph {et~al.}(2020)\citenamefont {Oh},
  \citenamefont {Gye},\ and\ \citenamefont {Yeom}}]{Oh2020}%
  \BibitemOpen
  \bibfield  {author} {\bibinfo {author} {\bibfnamefont {E.}~\bibnamefont
  {Oh}}, \bibinfo {author} {\bibfnamefont {G.}~\bibnamefont {Gye}},\ and\
  \bibinfo {author} {\bibfnamefont {H.~W.}\ \bibnamefont {Yeom}},\ }\bibfield
  {title} {\bibinfo {title} {Defect-selective charge-density-wave condensation
  in {2\textit{H}}--{NbSe}$_2$},\ }\href
  {https://doi.org/10.1103/physrevlett.125.036804} {\bibfield  {journal}
  {\bibinfo  {journal} {Phys. Rev. Lett.}\ }\textbf {\bibinfo {volume} {125}},\
  \bibinfo {pages} {036804} (\bibinfo {year} {2020})}\BibitemShut {NoStop}%
\bibitem [{\citenamefont {McMillan}(1976)}]{McMillan1976}%
  \BibitemOpen
  \bibfield  {author} {\bibinfo {author} {\bibfnamefont {W.~L.}\ \bibnamefont
  {McMillan}},\ }\bibfield  {title} {\bibinfo {title} {Theory of
  discommensurations and the commensurate-incommensurate charge-density-wave
  phase transition},\ }\href {https://doi.org/10.1103/physrevb.14.1496}
  {\bibfield  {journal} {\bibinfo  {journal} {Phys. Rev. B}\ }\textbf {\bibinfo
  {volume} {14}},\ \bibinfo {pages} {1496} (\bibinfo {year}
  {1976})}\BibitemShut {NoStop}%
\bibitem [{\citenamefont {Liu}\ \emph {et~al.}(2021)\citenamefont {Liu},
  \citenamefont {Chong}, \citenamefont {Sharma},\ and\ \citenamefont
  {Davis}}]{Liu2021}%
  \BibitemOpen
  \bibfield  {author} {\bibinfo {author} {\bibfnamefont {X.}~\bibnamefont
  {Liu}}, \bibinfo {author} {\bibfnamefont {Y.~X.}\ \bibnamefont {Chong}},
  \bibinfo {author} {\bibfnamefont {R.}~\bibnamefont {Sharma}},\ and\ \bibinfo
  {author} {\bibfnamefont {J.~C.~S.}\ \bibnamefont {Davis}},\ }\bibfield
  {title} {\bibinfo {title} {Discovery of a {Cooper}-pair density wave state in
  a transition-metal dichalcogenide},\ }\href
  {https://doi.org/10.1126/science.abd4607} {\bibfield  {journal} {\bibinfo
  {journal} {Science}\ }\textbf {\bibinfo {volume} {372}},\ \bibinfo {pages}
  {1447} (\bibinfo {year} {2021})}\BibitemShut {NoStop}%
\bibitem [{\citenamefont {Sanna}\ \emph {et~al.}(2022)\citenamefont {Sanna},
  \citenamefont {Pellegrini}, \citenamefont {Liebhaber}, \citenamefont
  {Rossnagel}, \citenamefont {Franke},\ and\ \citenamefont
  {Gross}}]{Sanna2022}%
  \BibitemOpen
  \bibfield  {author} {\bibinfo {author} {\bibfnamefont {A.}~\bibnamefont
  {Sanna}}, \bibinfo {author} {\bibfnamefont {C.}~\bibnamefont {Pellegrini}},
  \bibinfo {author} {\bibfnamefont {E.}~\bibnamefont {Liebhaber}}, \bibinfo
  {author} {\bibfnamefont {K.}~\bibnamefont {Rossnagel}}, \bibinfo {author}
  {\bibfnamefont {K.~J.}\ \bibnamefont {Franke}},\ and\ \bibinfo {author}
  {\bibfnamefont {E.~K.~U.}\ \bibnamefont {Gross}},\ }\bibfield  {title}
  {\bibinfo {title} {Real-space anisotropy of the superconducting gap in the
  charge-density wave material {2H}--{NbSe}$_2$},\ }\href
  {https://doi.org/10.1038/s41535-021-00412-8} {\bibfield  {journal} {\bibinfo
  {journal} {npj Quantum Mater.}\ }\textbf {\bibinfo {volume} {7}},\ \bibinfo
  {pages} {6} (\bibinfo {year} {2022})}\BibitemShut {NoStop}%
\bibitem [{\citenamefont {Lian}\ \emph {et~al.}(2018)\citenamefont {Lian},
  \citenamefont {Si},\ and\ \citenamefont {Duan}}]{Lian2018}%
  \BibitemOpen
  \bibfield  {author} {\bibinfo {author} {\bibfnamefont {C.-S.}\ \bibnamefont
  {Lian}}, \bibinfo {author} {\bibfnamefont {C.}~\bibnamefont {Si}},\ and\
  \bibinfo {author} {\bibfnamefont {W.}~\bibnamefont {Duan}},\ }\bibfield
  {title} {\bibinfo {title} {Unveiling charge-density wave, superconductivity,
  and their competitive nature in two-dimensional {NbSe}$_2$},\ }\href
  {https://doi.org/10.1021/acs.nanolett.8b00237} {\bibfield  {journal}
  {\bibinfo  {journal} {Nano Lett.}\ }\textbf {\bibinfo {volume} {18}},\
  \bibinfo {pages} {2924} (\bibinfo {year} {2018})}\BibitemShut {NoStop}%
\bibitem [{\citenamefont {P{\'{a}}sztor}\ \emph {et~al.}(2019)\citenamefont
  {P{\'{a}}sztor}, \citenamefont {Scarfato}, \citenamefont {Spera},
  \citenamefont {Barreteau}, \citenamefont {Giannini},\ and\ \citenamefont
  {Renner}}]{Pasztor2019}%
  \BibitemOpen
  \bibfield  {author} {\bibinfo {author} {\bibfnamefont {{\'{A}}.}~\bibnamefont
  {P{\'{a}}sztor}}, \bibinfo {author} {\bibfnamefont {A.}~\bibnamefont
  {Scarfato}}, \bibinfo {author} {\bibfnamefont {M.}~\bibnamefont {Spera}},
  \bibinfo {author} {\bibfnamefont {C.}~\bibnamefont {Barreteau}}, \bibinfo
  {author} {\bibfnamefont {E.}~\bibnamefont {Giannini}},\ and\ \bibinfo
  {author} {\bibfnamefont {C.}~\bibnamefont {Renner}},\ }\bibfield  {title}
  {\bibinfo {title} {Holographic imaging of the complex charge density wave
  order parameter},\ }\href {https://doi.org/10.1103/physrevresearch.1.033114}
  {\bibfield  {journal} {\bibinfo  {journal} {Phys. Rev. Res.}\ }\textbf
  {\bibinfo {volume} {1}},\ \bibinfo {pages} {033114} (\bibinfo {year}
  {2019})}\BibitemShut {NoStop}%
\bibitem [{\citenamefont {Momma}\ and\ \citenamefont
  {Izumi}(2011)}]{Momma2011}%
  \BibitemOpen
  \bibfield  {author} {\bibinfo {author} {\bibfnamefont {K.}~\bibnamefont
  {Momma}}\ and\ \bibinfo {author} {\bibfnamefont {F.}~\bibnamefont {Izumi}},\
  }\bibfield  {title} {\bibinfo {title} {{VESTA}3 for three-dimensional
  visualization of crystal, volumetric and morphology data},\ }\href
  {https://doi.org/10.1107/s0021889811038970} {\bibfield  {journal} {\bibinfo
  {journal} {J. Appl. Crystallogr.}\ }\textbf {\bibinfo {volume} {44}},\
  \bibinfo {pages} {1272} (\bibinfo {year} {2011})}\BibitemShut {NoStop}%
\bibitem [{Sup()}]{Supplement}%
  \BibitemOpen
  \href@noop {} {}\bibinfo {note} {See Supplemental Material at URL, which
  includes Refs. \cite{Bray2002,Vogelgesang2017}.}\BibitemShut {Stop}%
\bibitem [{\citenamefont {Lawler}\ \emph {et~al.}(2010)\citenamefont {Lawler},
  \citenamefont {Fujita}, \citenamefont {Lee}, \citenamefont {Schmidt},
  \citenamefont {Kohsaka}, \citenamefont {Kim}, \citenamefont {Eisaki},
  \citenamefont {Uchida}, \citenamefont {Davis}, \citenamefont {Sethna},\ and\
  \citenamefont {Kim}}]{Lawler2010}%
  \BibitemOpen
  \bibfield  {author} {\bibinfo {author} {\bibfnamefont {M.~J.}\ \bibnamefont
  {Lawler}}, \bibinfo {author} {\bibfnamefont {K.}~\bibnamefont {Fujita}},
  \bibinfo {author} {\bibfnamefont {J.}~\bibnamefont {Lee}}, \bibinfo {author}
  {\bibfnamefont {A.~R.}\ \bibnamefont {Schmidt}}, \bibinfo {author}
  {\bibfnamefont {Y.}~\bibnamefont {Kohsaka}}, \bibinfo {author} {\bibfnamefont
  {C.~K.}\ \bibnamefont {Kim}}, \bibinfo {author} {\bibfnamefont
  {H.}~\bibnamefont {Eisaki}}, \bibinfo {author} {\bibfnamefont
  {S.}~\bibnamefont {Uchida}}, \bibinfo {author} {\bibfnamefont {J.~C.}\
  \bibnamefont {Davis}}, \bibinfo {author} {\bibfnamefont {J.~P.}\ \bibnamefont
  {Sethna}},\ and\ \bibinfo {author} {\bibfnamefont {E.-A.}\ \bibnamefont
  {Kim}},\ }\bibfield  {title} {\bibinfo {title} {Intra-unit-cell electronic
  nematicity of the high-tc copper-oxide pseudogap states},\ }\href
  {https://doi.org/10.1038/nature09169} {\bibfield  {journal} {\bibinfo
  {journal} {Nature (London)}\ }\textbf {\bibinfo {volume} {466}},\ \bibinfo
  {pages} {347} (\bibinfo {year} {2010})}\BibitemShut {NoStop}%
\bibitem [{\citenamefont {Walker}\ and\ \citenamefont
  {Jacobs}(1981)}]{Walker1981}%
  \BibitemOpen
  \bibfield  {author} {\bibinfo {author} {\bibfnamefont {M.~B.}\ \bibnamefont
  {Walker}}\ and\ \bibinfo {author} {\bibfnamefont {A.~E.}\ \bibnamefont
  {Jacobs}},\ }\bibfield  {title} {\bibinfo {title} {Distinct commensurate
  charge-density-wave phases in the {2\textit{H}}--{NbSe}$_2$},\ }\href
  {https://doi.org/10.1103/physrevb.24.6770} {\bibfield  {journal} {\bibinfo
  {journal} {Phys. Rev. B}\ }\textbf {\bibinfo {volume} {24}},\ \bibinfo
  {pages} {6770} (\bibinfo {year} {1981})}\BibitemShut {NoStop}%
\bibitem [{\citenamefont {Nakanishi}\ and\ \citenamefont
  {Shiba}(1983)}]{Nakanishi1983}%
  \BibitemOpen
  \bibfield  {author} {\bibinfo {author} {\bibfnamefont {K.}~\bibnamefont
  {Nakanishi}}\ and\ \bibinfo {author} {\bibfnamefont {H.}~\bibnamefont
  {Shiba}},\ }\bibfield  {title} {\bibinfo {title} {Theory of
  discommensurations and re-entrant lock-in transition in the
  charge-density-wave state of {2H}-{TaSe}$_2$},\ }\href
  {https://doi.org/10.1143/jpsj.52.1278} {\bibfield  {journal} {\bibinfo
  {journal} {J. Phys. Soc. Jpn.}\ }\textbf {\bibinfo {volume} {52}},\ \bibinfo
  {pages} {1278} (\bibinfo {year} {1983})}\BibitemShut {NoStop}%
\bibitem [{\citenamefont {Shiba}\ and\ \citenamefont
  {Nakanishi}(1986)}]{Nakanishi1986}%
  \BibitemOpen
  \bibfield  {author} {\bibinfo {author} {\bibfnamefont {H.}~\bibnamefont
  {Shiba}}\ and\ \bibinfo {author} {\bibfnamefont {K.}~\bibnamefont
  {Nakanishi}},\ }\bibinfo {title} {Phenomenological {Landau} theory of charge
  density wave phase transitions in layered compounds},\ in\ \href@noop {}
  {\emph {\bibinfo {booktitle} {Structural Phase Transitions in Layered
  Transition-metal Compounds}}},\ \bibinfo {editor} {edited by\ \bibinfo
  {editor} {\bibfnamefont {K.}~\bibnamefont {Motizuki}}}\ (\bibinfo
  {publisher} {Springer}, {Dordrecht},\ \bibinfo {year} {1986})\ pp.\ \bibinfo {pages}
  {175--266}\BibitemShut {NoStop}%
\bibitem [{\citenamefont {Giambattista}\ \emph {et~al.}(1988)\citenamefont
  {Giambattista}, \citenamefont {Johnson}, \citenamefont {Coleman},
  \citenamefont {Drake},\ and\ \citenamefont {Hansma}}]{Giambattista1988}%
  \BibitemOpen
  \bibfield  {author} {\bibinfo {author} {\bibfnamefont {B.}~\bibnamefont
  {Giambattista}}, \bibinfo {author} {\bibfnamefont {A.}~\bibnamefont
  {Johnson}}, \bibinfo {author} {\bibfnamefont {R.~V.}\ \bibnamefont
  {Coleman}}, \bibinfo {author} {\bibfnamefont {B.}~\bibnamefont {Drake}},\
  and\ \bibinfo {author} {\bibfnamefont {P.~K.}\ \bibnamefont {Hansma}},\
  }\bibfield  {title} {\bibinfo {title} {Charge-density waves observed at 4.2 {K}
  by scanning-tunneling microscopy},\ }\href
  {https://doi.org/10.1103/physrevb.37.2741} {\bibfield  {journal} {\bibinfo
  {journal} {Phys. Rev. B}\ }\textbf {\bibinfo {volume} {37}},\ \bibinfo
  {pages} {2741} (\bibinfo {year} {1988})}\BibitemShut {NoStop}%
\bibitem [{\citenamefont {Nakanishi}(1979)}]{Nakanishi1979}%
  \BibitemOpen
  \bibfield  {author} {\bibinfo {author} {\bibfnamefont {K.}~\bibnamefont
  {Nakanishi}},\ }\bibfield  {title} {\bibinfo {title} {Effects of impurity
  pinning on commensurate charge-density-wave state and
  incommensurate-commensurate transition},\ }\href
  {https://doi.org/10.1143/jpsj.46.1434} {\bibfield  {journal} {\bibinfo
  {journal} {J. Phys. Soc. Jpn.}\ }\textbf {\bibinfo {volume} {46}},\ \bibinfo
  {pages} {1434} (\bibinfo {year} {1979})}\BibitemShut {NoStop}%
\bibitem [{\citenamefont {Rice}\ \emph {et~al.}(1981)\citenamefont {Rice},
  \citenamefont {Whitehouse},\ and\ \citenamefont {Littlewood}}]{Rice1981}%
  \BibitemOpen
  \bibfield  {author} {\bibinfo {author} {\bibfnamefont {T.~M.}\ \bibnamefont
  {Rice}}, \bibinfo {author} {\bibfnamefont {S.}~\bibnamefont {Whitehouse}},\
  and\ \bibinfo {author} {\bibfnamefont {P.}~\bibnamefont {Littlewood}},\
  }\bibfield  {title} {\bibinfo {title} {Impurity pinning of discommensurations
  in charge-density waves},\ }\href {https://doi.org/10.1103/physrevb.24.2751}
  {\bibfield  {journal} {\bibinfo  {journal} {Phys. Rev. B}\ }\textbf {\bibinfo
  {volume} {24}},\ \bibinfo {pages} {2751} (\bibinfo {year}
  {1981})}\BibitemShut {NoStop}%
\bibitem [{\citenamefont {Arguello}\ \emph {et~al.}(2014)\citenamefont
  {Arguello}, \citenamefont {Chockalingam}, \citenamefont {Rosenthal},
  \citenamefont {Zhao}, \citenamefont {Guti{\'{e}}rrez}, \citenamefont {Kang},
  \citenamefont {Chung}, \citenamefont {Fernandes}, \citenamefont {Jia},
  \citenamefont {Millis}, \citenamefont {Cava},\ and\ \citenamefont
  {Pasupathy}}]{Arguello2014}%
  \BibitemOpen
  \bibfield  {author} {\bibinfo {author} {\bibfnamefont {C.~J.}\ \bibnamefont
  {Arguello}}, \bibinfo {author} {\bibfnamefont {S.~P.}\ \bibnamefont
  {Chockalingam}}, \bibinfo {author} {\bibfnamefont {E.~P.}\ \bibnamefont
  {Rosenthal}}, \bibinfo {author} {\bibfnamefont {L.}~\bibnamefont {Zhao}},
  \bibinfo {author} {\bibfnamefont {C.}~\bibnamefont {Guti{\'{e}}rrez}},
  \bibinfo {author} {\bibfnamefont {J.~H.}\ \bibnamefont {Kang}}, \bibinfo
  {author} {\bibfnamefont {W.~C.}\ \bibnamefont {Chung}}, \bibinfo {author}
  {\bibfnamefont {R.~M.}\ \bibnamefont {Fernandes}}, \bibinfo {author}
  {\bibfnamefont {S.}~\bibnamefont {Jia}}, \bibinfo {author} {\bibfnamefont
  {A.~J.}\ \bibnamefont {Millis}}, \bibinfo {author} {\bibfnamefont {R.~J.}\
  \bibnamefont {Cava}},\ and\ \bibinfo {author} {\bibfnamefont {A.~N.}\
  \bibnamefont {Pasupathy}},\ }\bibfield  {title} {\bibinfo {title}
  {Visualizing the charge density wave transition in {2\textit{H}}--{NbSe}$_2$
  in real space},\ }\href {https://doi.org/10.1103/physrevb.89.235115}
  {\bibfield  {journal} {\bibinfo  {journal} {Phys. Rev. B}\ }\textbf {\bibinfo
  {volume} {89}},\ \bibinfo {pages} {235115} (\bibinfo {year}
  {2014})}\BibitemShut {NoStop}%
\bibitem [{\citenamefont {Kibble}(2007)}]{Kibble2007}%
  \BibitemOpen
  \bibfield  {author} {\bibinfo {author} {\bibfnamefont {T.}~\bibnamefont
  {Kibble}},\ }\bibfield  {title} {\bibinfo {title} {Phase-transition dynamics
  in the lab and the universe},\ }\href {https://doi.org/10.1063/1.2784684}
  {\bibfield  {journal} {\bibinfo  {journal} {Phys. Today}\ }\textbf {\bibinfo
  {volume} {60}},\ \bibinfo {pages} {47} (\bibinfo {year} {2007})}\BibitemShut
  {NoStop}%
\bibitem [{\citenamefont {Okamoto}\ \emph {et~al.}(2015)\citenamefont
  {Okamoto}, \citenamefont {Arguello}, \citenamefont {Rosenthal}, \citenamefont
  {Pasupathy},\ and\ \citenamefont {Millis}}]{Okamoto2015}%
  \BibitemOpen
  \bibfield  {author} {\bibinfo {author} {\bibfnamefont {J.~I.}~\bibnamefont
  {Okamoto}}, \bibinfo {author} {\bibfnamefont {C.~J.}\ \bibnamefont
  {Arguello}}, \bibinfo {author} {\bibfnamefont {E.~P.}\ \bibnamefont
  {Rosenthal}}, \bibinfo {author} {\bibfnamefont {A.~N.}\ \bibnamefont
  {Pasupathy}},\ and\ \bibinfo {author} {\bibfnamefont {A.~J.}\ \bibnamefont
  {Millis}},\ }\bibfield  {title} {\bibinfo {title} {Experimental evidence for
  a {Bragg} glass density wave phase in a transition-metal dichalcogenide},\
  }\href {https://doi.org/10.1103/physrevlett.114.026802} {\bibfield  {journal}
  {\bibinfo  {journal} {Phys. Rev. Lett.}\ }\textbf {\bibinfo {volume} {114}},\
  \bibinfo {pages} {026802} (\bibinfo {year} {2015})}\BibitemShut {NoStop}%
\bibitem [{\citenamefont {Bray}(2002)}]{Bray2002}%
  \BibitemOpen
  \bibfield  {author} {\bibinfo {author} {\bibfnamefont {A.~J.}\ \bibnamefont
  {Bray}},\ }\bibfield  {title} {\bibinfo {title} {Theory of phase-ordering
  kinetics},\ }\href {https://doi.org/10.1080/00018730110117433} {\bibfield
  {journal} {\bibinfo  {journal} {Adv. Phys.}\ }\textbf {\bibinfo {volume}
  {51}},\ \bibinfo {pages} {481} (\bibinfo {year} {2002})}\BibitemShut
  {NoStop}%
\bibitem [{\citenamefont {Vogelgesang}\ \emph {et~al.}(2017)\citenamefont
  {Vogelgesang}, \citenamefont {Storeck}, \citenamefont {Horstmann},
  \citenamefont {Diekmann}, \citenamefont {Sivis}, \citenamefont {Schramm},
  \citenamefont {Rossnagel}, \citenamefont {Sch{\"a}fer},\ and\ \citenamefont
  {Roper}}]{Vogelgesang2017}%
  \BibitemOpen
  \bibfield  {author} {\bibinfo {author} {\bibfnamefont {S.}~\bibnamefont
  {Vogelgesang}}, \bibinfo {author} {\bibfnamefont {G.}~\bibnamefont
  {Storeck}}, \bibinfo {author} {\bibfnamefont {J.~G.}\ \bibnamefont
  {Horstmann}}, \bibinfo {author} {\bibfnamefont {T.}~\bibnamefont {Diekmann}},
  \bibinfo {author} {\bibfnamefont {M.}~\bibnamefont {Sivis}}, \bibinfo
  {author} {\bibfnamefont {S.}~\bibnamefont {Schramm}}, \bibinfo {author}
  {\bibfnamefont {K.}~\bibnamefont {Rossnagel}}, \bibinfo {author}
  {\bibfnamefont {S.}~\bibnamefont {Sch{\"a}fer}},\ and\ \bibinfo {author}
  {\bibfnamefont {C.}~\bibnamefont {Roper}},\ }\bibfield  {title} {\bibinfo
  {title} {Phase ordering of charge density waves traced by ultrafast
  low-energy electron diffraction},\ }\href {https://doi.org/10.1038/nphys4309}
  {\bibfield  {journal} {\bibinfo  {journal} {Nat. Phys.}\ }\textbf {\bibinfo
  {volume} {14}},\ \bibinfo {pages} {184} (\bibinfo {year} {2017})}\BibitemShut
  {NoStop}%
\end{thebibliography}

%

\end{document}


\title{ Supplemental Material for \\ 
Visualization of Alternating Triangular Domains of Charge Density Waves in 2\textit{H}--NbSe$_2$ by Scanning Tunneling Microscopy }

\author{Shunsuke Yoshizawa}\email{YOSHIZAWA.Shunsuke@nims.go.jp}\affiliation{Center for Basic Research on Materials, National Institute for Materials Science, 1-2-1 Sengen, Tsukuba, Ibaraki 305-0047, Japan}
\author{Keisuke Sagisaka}\affiliation{Center for Basic Research on Materials, National Institute for Materials Science, 1-2-1 Sengen, Tsukuba, Ibaraki 305-0047, Japan}
\author{Hideaki Sakata}\affiliation{Department of Physics, Tokyo University of Science, 1-3 Kagurazaka, Shinjuku, Tokyo 162-8601, Japan}

\date{\today}

\begin{abstract}~~
\end{abstract}

\maketitle

\section{Possible combinations of indices and corresponding commensurate CDW structures}

\begin{figure*}
\includegraphics[width=\linewidth]{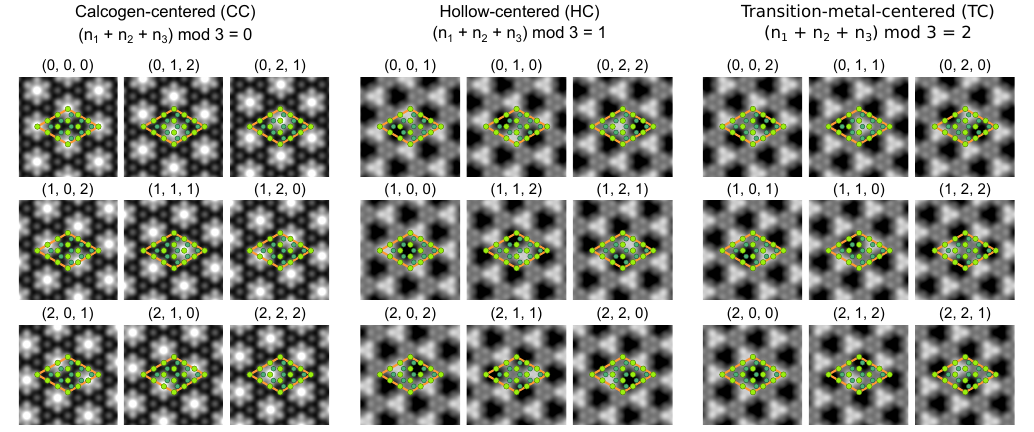}
\caption{\label{figs-sim}
Simulated topographic images of commensurate CDW structures labeled $(n_1, n_2, n_3)$. 
The large dots in light green and small dots in green represent the Se and Nb sites, respectively. 
The orange rhombus in each image indicates the $3 \times 3$ cell of the CDW and is fixed to the atomic lattice to show the relative position of the CDW maxima.
}\end{figure*}

In the main text, we model the topography of scanning tunneling microscopy (STM) by the sum of the local density of states (LDOS) from the crystal lattice and that from the charge density wave (CDW).
It is expressed as
\begin{equation}
    \rho^{(n_1, n_2, n_3)}(\mathbf{r}) = \rho_\mathrm{Se}(\mathbf{r}) + c \rho_\mathrm{CDW}^{(n_1, n_2, n_3)}(\mathbf{r}),
\end{equation}
The coefficient $c > 0$ is used to adjust the image contrast.
We have selected $c \sim 2.5$ for our simulations.
The LDOS from the crystal lattice, which reflects the locations of Se sites, is given by
\begin{equation}
     \rho_\mathrm{Se}(\mathbf{r}) = \operatorname{Re} \sum_{j=1}^3 \exp(i\mathbf{b}_j\cdot\mathbf{r}),
\label{eq:rhose}\end{equation}
while the LDOS from the CDW is given by
\begin{equation}
    \rho_\mathrm{CDW}^{(n_1, n_2, n_3)}(\mathbf{r}) = \operatorname{Re} \sum_{j=1}^3 \exp\left[i\left(\mathbf{Q}_j\cdot\mathbf{r} - \frac{2\pi}{3}n_j + \varphi_0 \right)\right]
\label{eq:rhocdwn}\end{equation}
with commensurate wavevectors $\mathbf{Q}_j = (1/3)\mathbf{b}_j$.
The values of $n_j$ ($j = 1, 2, 3)$ control the positions of the CDW modulations relative to the crystal lattice.
We can set a small phase offset $\varphi_0$ to adjust the appearance of simulated images, but we omit it for simplicity ($\varphi_0 = 0$). 
In commensurate CDW domains, $n_j$ takes integer values, and the type of commensurate structure depends on $\lambda \equiv (n_1 + n_2 + n_3) \mod 3$.
Specifically, when $\lambda = 0$, $\rho^{(n_1, n_2, n_3)}(\mathbf{r})$ reproduces the chalcogen-centered (CC) structure, where CDW maxima are located at chalcogen sites. 
When $\lambda = 1$, $\rho^{(n_1, n_2, n_3)}(\mathbf{r})$ reproduces the hollow-centered (HC) structure, where CDW maxima are at hollow sites.
When $\lambda = 2$, the CDW maxima are at transition-metal sites, and the corresponding structure could be called the transition-metal-centered (TC) structure.
However, this type of structure can be only accidentally found at the vertices of the CDW domains.
The absence of the TC structure is probably beneficial in lowering the total energy, since density functional theory (DFT) calculations show that the formation energy of the TC structure is higher than that of the CC and HC structures \cite{Gye2019}.
The unit cell of the commensurate CDW contains nine chalcogen sites, nine hollow sites, and nine transition metal sites.
Therefore, each of the CC, HC, and TC structures has nine variants that differ in phase and are distinguished by the label $(n_1, n_2, n_3)$.
All of these variants are illustrated in Fig. \ref{figs-sim}.

\section{Full procedure to determine CDW domains from topographic image}

\subsection{Summary of Lawler--Fujita method}

We summarize the method of extracting the displacement field of a periodic structure from a topographic image proposed by Lawler and Fujita \textit{et al.} \cite{Lawler2010}.
This method is essentially a lock-in technique with respect to the spatial coordinates.
Let $z(\mathbf{r})$ be a topographic image that contains a periodic structure with a wavevector $\mathbf{q}$.
The periodic structure may have a small strain, which is represented by a displacement field $\mathbf{u}(\mathbf{r})$.
We can write the resulting topographic image as
\begin{equation}
    z(\mathbf{r}) = z_0 \cos[ \mathbf{q}\cdot(\mathbf{r} - \mathbf{u}(\mathbf{r}))] \mathrm{+ ~other ~periodic ~components}.
\end{equation}
The key quantity in the Lawler--Fujita method is
\begin{equation}
    C_{\mathbf{q}}(\mathbf{r}) \equiv \sum_{\mathbf{r}'} z(\mathbf{r}') \exp(-i\mathbf{q}\cdot\mathbf{r}') w(\mathbf{r} - \mathbf{r}'),
\end{equation}
where $w(\mathbf{r})$ is a normalized two-dimensional Gaussian function given by
\begin{equation}
    w(\mathbf{r}) = \frac{1}{2\pi\sigma^2}\exp[-|\mathbf{r}|^2/(2\sigma^2)].
\end{equation} 
Since $C_{\mathbf{q}}(\mathbf{r})$ is the convolution of $z(\mathbf{r})\exp(-i\mathbf{q}\cdot\mathbf{r})$ and $w(\mathbf{r})$, it can be computed efficiently by using the fast Fourier transform.
If $\sigma$ is smaller than the length scale of the $\mathbf{u}(\mathbf{r})$ variation, we can approximate $C_{\mathbf{q}}(\mathbf{r})$ as
\begin{equation}
    C_{\mathbf{q}}(\mathbf{r}) \simeq (z_0/2) \exp[-i\mathbf{q}\cdot\mathbf{u}(\mathbf{r})].
\end{equation}
Hence, we can obtain information about $\mathbf{u}(\mathbf{r})$ from the phase factor
\begin{equation}
    \varphi_\mathbf{q}(\mathbf{r}) \equiv \arg[ C_{\mathbf{q}}(\mathbf{r}) ].
\end{equation}
To remove the discreteness of $\varphi_\mathbf{q}(\mathbf{r})$ at $\pm\pi$, we must unwrap the phase to obtain a continuous function.
The resulting quantity provides the displacement field $\mathbf{u}(\mathbf{r})$ projected in the direction of $\mathbf{q}$,
\begin{equation}
    \mathbf{q}\cdot\mathbf{u}(\mathbf{r}) = -\operatorname{unwrap}[ \varphi_\mathbf{q}(\mathbf{r}) ].
\end{equation}
We implemented the above procedure using the Python language. For two-dimensional phase unwrapping, we simply called the \texttt{skimage.restoration.unwrap\_phase} function in the scikit-image library.

\subsection{Determination of intrinsic displacement field of CDW}

\begin{figure}
    \includegraphics[width=\linewidth]{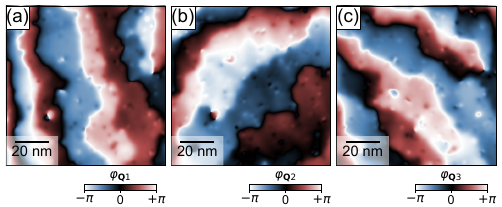}
    \caption{\label{figs-cdwphase}
    (a)--(c) Images of $\varphi_{\mathbf{Q}_1}(\mathbf{r})$, $\varphi_{\mathbf{Q}_2}(\mathbf{r})$, and $\varphi_{\mathbf{Q}_3}(\mathbf{r})$ obtained from the topographic image in Fig. 2(a) of the main text.
}\end{figure}
    
\begin{figure}
    \includegraphics[width=\linewidth]{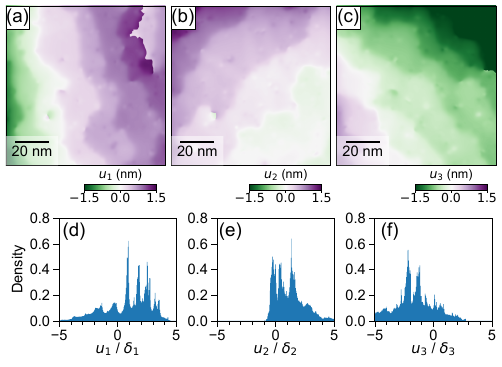}
    \caption{\label{figs-cdwdisp}
    (a)--(c) Images of $u_1(\mathbf{r})$, $u_2(\mathbf{r})$, and $u_3(\mathbf{r})$.
    (d)--(f) Histograms of $u_1(\mathbf{r})$, $u_2(\mathbf{r})$, and $u_3(\mathbf{r})$.
    The transverse axis is normalized by the interplanar spacing ($\delta_j \equiv 2\pi/|\mathbf{b}_j|$) to facilitate comparison with the histograms of $n_1(\mathbf{r})$, $n_2(\mathbf{r})$, and $n_3(\mathbf{r})$ in Fig. 3 in the main text. 
    The vertical axis represents frequency density and is normalized so that the total area is 1.
}\end{figure}
    
Figures \ref{figs-cdwphase}(a)--\ref{figs-cdwphase}(c) shows the images of $\varphi_{\mathbf{Q}_j}(\mathbf{r})$ ($j = 1, 2, 3$) obtained by the Lawler--Fujita method from the topographic image in Fig. 2(a) of the main text. 
By unwrapping these phase images, we obtain the displacement field of the $j$-th CDW component projected to the direction of $\mathbf{Q}_j$, which is written as
\begin{equation}
    u_j(\mathbf{r}) \equiv \mathbf{u}_j(\mathbf{r})\cdot\mathbf{Q}_j/|\mathbf{Q}_j| = -\operatorname{unwrap}[ \varphi_{\mathbf{Q}_j}(\mathbf{r}) ] / |\mathbf{Q}_j|.
\end{equation}
The resulting images are shown in Figs. \ref{figs-cdwdisp}(a)--\ref{figs-cdwdisp}(c).
Note that this displacement field is the sum of the intrinsic CDW displacements and the extrinsic image deformation due to the creep of the piezoelectric scanner of the STM head.
Owing to the latter effect, the histograms of $u_1(\mathbf{r})$, $u_2(\mathbf{r})$, and $u_3(\mathbf{r})$ do not exhibit sharp peaks and contains a broad background, as shown in Figs. \ref{figs-cdwdisp}(d)--\ref{figs-cdwdisp}(f).

\begin{figure}
    \includegraphics[width=\linewidth]{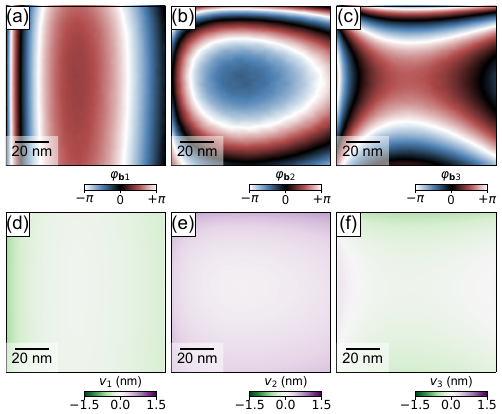}
    \caption{\label{figs-latphase}
    (a)--(c) Images of $\varphi_{\mathbf{b}_1}(\mathbf{r})$, $\varphi_{\mathbf{b}_2}(\mathbf{r})$, and $\varphi_{\mathbf{b}_3}(\mathbf{r})$.
    (d)--(f) Images of $v_1(\mathbf{r})$, $v_2(\mathbf{r})$, and $v_3(\mathbf{r})$.
}\end{figure}

To eliminate the extrinsic image deformation, we measure the displacement field of the crystal lattice, denoted by $\mathbf{v}(\mathbf{r})$.
Since the crystal lattice should have perfect periodicity, the apparent displacement field of the crystal lattice in the topographic image is attributed purely to the extrinsic image deformation.
By applying the Lawler--Fujita method to the periodicity of the crystal lattice, $\mathbf{b}_j$, we obtain the phase images, $\varphi_{\mathbf{b}_j}(\mathbf{r})$ [Figs. \ref{figs-latphase}(a)--\ref{figs-latphase}(c)], and the displacement fields, $v_j(\mathbf{r}) \equiv \mathbf{v}(\mathbf{r})\cdot\mathbf{b}_j/|\mathbf{b}_j|$ [Figs. \ref{figs-latphase}(d)--\ref{figs-latphase}(f)].
The intrinsic displacement of the CDW can then be determined as 
\begin{equation}
    \delta u_j(\mathbf{r}) = u_j(\mathbf{r}) - v_j(\mathbf{r}).
\end{equation}

This argument, which assumes the perfect periodicity of the crystal lattice, may seem strange at first glance. 
For example, the CDW transition would be accompanied by lattice deformation at the same period as the CDW. 
Also, atomic defects could induce a local lattice distortion around them.
However, the displacement field estimated by the Lawler--Fujita method is smoothed by a Gaussian window function characterized by $\sigma$. 
If $\sigma$ is set sufficiently large, the lattice distortions mentioned above average out, because they are periodic or quite local. 
The insensitivity to local lattice distortions is verified by the smoothness of $\varphi_{\mathbf{b}_j}(\mathbf{r})$ and $v_j(\mathbf{r})$ shown in Fig. \ref{figs-latphase}.

One might be concerned about the influence of the harmonics of the CDW on the determination of the displacement field of the crystal lattice, since the locations of the third-order Fourier peaks of the CDW coincide with those of the first-order peaks of the crystal lattice [Figs. 2(b) and 2(c) in the main text]. 
Such an influence can be neglected in the present case, because the contribution of the crystal lattice is dominant in the Bragg peaks. 
This is confirmed by the fact that the displacement fields determined for the periodicities $\mathbf{b}_1$--$\mathbf{b}_3$ are smooth functions, without any recognizable features related to the domain structure of the CDW.
See $\varphi_{\mathbf{b}_j}(\mathbf{r})$ and $v_j(\mathbf{r})$ shown in Fig. \ref{figs-latphase}.

\subsection{Determination of $(n_1, n_2, n_3)$ and visualization of domain structure}

\begin{figure}
    \includegraphics[width=\linewidth]{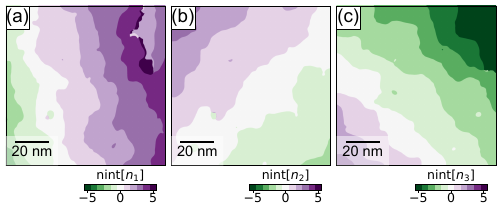}
    \caption{\label{figs-nint}
    (a)--(c) Images of $\mathrm{nint}[n_1(\mathbf{r})]$, $\mathrm{nint}[n_2(\mathbf{r})]$, and $\mathrm{nint}[n_3(\mathbf{r})]$.
}\end{figure}

\begin{figure}
    \includegraphics[width=\linewidth]{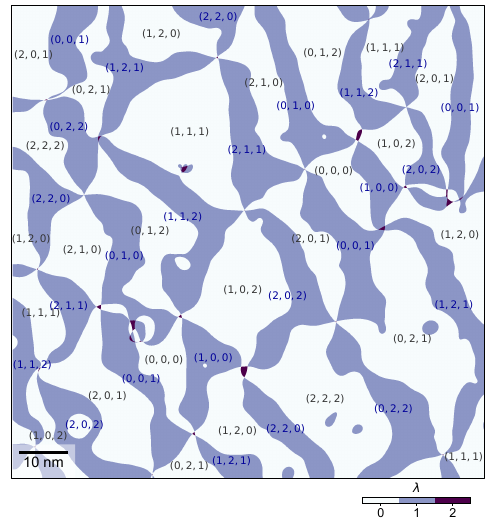}
    \caption{\label{figs-redidx}
    (a)--(c)Image of $\lambda(\mathbf{r})$ of the same region studied in the main text. 
    The CC and HC domains are depicted as white and blue areas, respectively.
     Each domain is labeled by reduced indices given by $(n_1 ~\mathrm{mod}~ 3, n_2 ~\mathrm{mod}~ 3, n_3 ~\mathrm{mod}~ 3)$.
}\end{figure}

\begin{figure*}
    \includegraphics[width=0.77\linewidth]{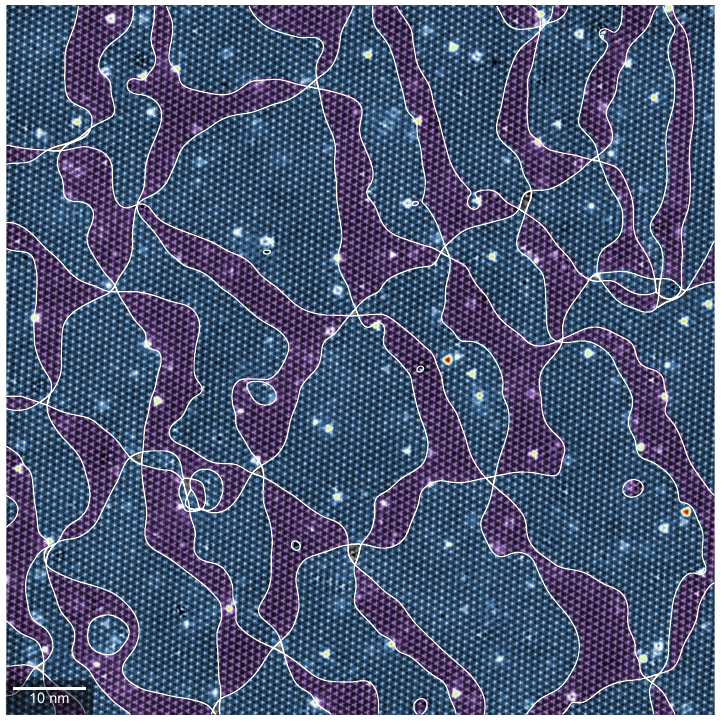}
    \caption{\label{figs-composit}
    Topographic image plotted with different colors corresponding to the domain type.
    The blue and purple areas correspond to the CC and HC regions, respectively.
    The original $2048 \times 2048$ pixels image was resampled to $1024 \times 1024$ pixels to reduce the file size.
}\end{figure*}

By relating the intrinsic displacement field $\delta u_j$ to the $(2 \pi/ 3)n_j$ in Eq. (\ref{eq:rhocdwn}) and using $|\mathbf{Q}_j| = (1/3)|\mathbf{b}_j|$, we see that $n_j$ can be calculated as
\begin{equation}
    n_j(\mathbf{r}) = \frac{|\mathbf{b}_j|}{2\pi} \delta u_j(\mathbf{r}).
\end{equation}
Note that $|\mathbf{b}_j|/(2\pi)$ is the inverse of the interplanar spacing measured in the direction of $\mathbf{b}_j$.
Figures 3(a)--3(c) of the main text display the obtained $n_1(\mathbf{r})$, $n_2(\mathbf{r})$, and $n_3(\mathbf{r})$.
The corresponding histograms in Figs. 3(d)--3(f) exhibit sharp, equally spaced peaks at integers in contrast to the histograms obtained before subtracting the extrinsic deformation [Figs. \ref{figs-cdwdisp}(d)--\ref{figs-cdwdisp}(f)].
To visualize the domain structure, we define
\begin{equation}
    \lambda(\mathbf{r}) \equiv \left\{ \sum_{j=1}^3 \mathrm{nint}[n_j(\mathbf{r})] \right\} ~\mathrm{mod}~ 3,
\end{equation}
where $\mathrm{nint}$ denotes the nearest integer.
The images of $\mathrm{nint}[n_1(\mathbf{r})]$, $\mathrm{nint}[n_2(\mathbf{r})]$, and $\mathrm{nint}[n_3(\mathbf{r})]$ are shown in Figs. \ref{figs-nint}(a)--\ref{figs-nint}(c).
The mapping of $\lambda(\mathbf{r})$, together with the triplet labels $(n_1, n_2, n_3)$, is shown in Fig. 3(g) of the main text.
Note that integers congruent modulo 3 can be considered equivalent as the label of the domains. For example, $(1, 0, -1)$ is equivalent to $(1, 0, 2)$, and the corresponding CDW images are identical. 
Considering this property, all of the CC and HC structures listed in Fig. \ref{figs-sim} can be assigned to the observed domains.
To illustrate this, in Fig. \ref{figs-redidx} we present the same domain structure as in Fig. 3(g) but labeled with reduced indices given by $(n_1 ~\mathrm{mod}~ 3, n_2 ~\mathrm{mod}~ 3, n_3 ~\mathrm{mod}~ 3)$.

In Fig. \ref{figs-composit}, we present an atomically resolved topographic image plotted with different colors corresponding to values of $\lambda$ to highlight the successful identification of the domain structure.

\subsection{Choice of the width of the Gaussian function}

\begin{figure}
    \includegraphics[width=\linewidth]{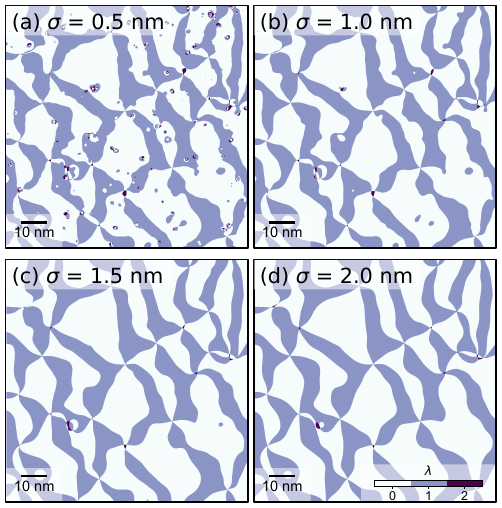}
    \caption{\label{figs-sigma}
    (a)--(c) Domain images obtained with $\sigma$ values of (a) 0.5 nm, (b) 1.0 nm, (c) 1.5 nm, and (d) 2.0 nm.
}\end{figure}

\begin{figure}
    \includegraphics[width=\linewidth]{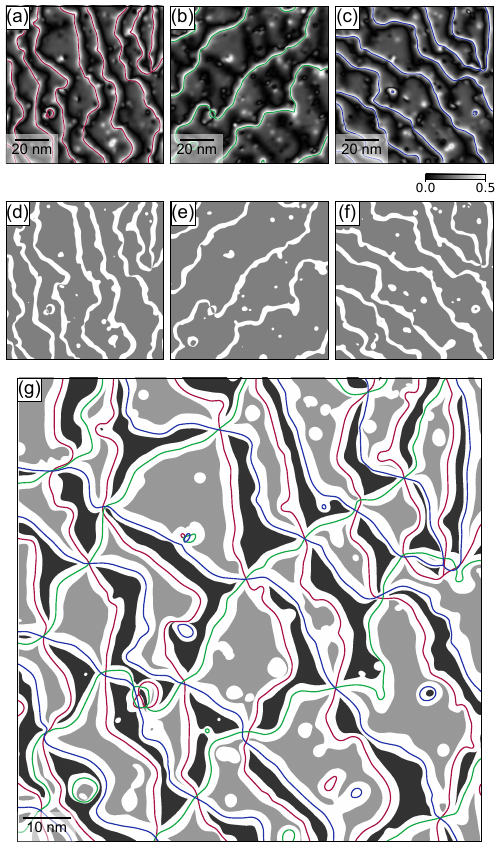}
    \caption{\label{figs-width}
    (a)--(c) $\delta n_1(\mathbf{r})$, $\delta n_2(\mathbf{r})$, and $\delta n_3(\mathbf{r})$.
    The colored curves show the contours of $n_j(\mathbf{r})$ at half-integer values, illustrating discommensurations.
    (d)--(f) Maps of discommensurate regions. 
    Areas where $\delta n_j(\mathbf{r}) > 0.25$ ($ \leq 0.25$) are in white (gray).
    (g) Distribution of CC (gray region) and HC (black region) domains with discommensurate regions (white region) removed. 
    The discommensurations determined from $n_1(\mathbf{r})$, $n_2(\mathbf{r})$, and $n_3(\mathbf{r})$ are depicted by curves in red, green, and blue, respectively.
}\end{figure}

The value of $\sigma$ was optimized by examining the balance between the resolution and smoothness. 
When we set $\sigma = 0.5$ nm, the detection of displacement fields is severely affected by the defects on the surface, as shown in Fig. \ref{figs-sigma}(a).
By setting larger $\sigma$ values, the displacement field becomes smoother [Figs. \ref{figs-sigma}(b)--\ref{figs-sigma}(d)].
On the other hand, Table \ref{tabS1} indicates that the width of discommensurations is slightly overestimated when $\sigma = 1.5$ or 2 nm (see Section S3).
We chose $\sigma = 1$ nm for our analysis, because the influence of surface defects is not significant and the estimated width of the discommensurate region remains to be as small as 3 nm.

\section{Analysis of discommensurate region}

\begin{table}
    \centering
    \begin{tabular}{c|cccc}
        $\sigma$ (nm) & $d_1$ (nm) & $d_2$ (nm) & $d_3$ (nm) & $\bar{d}$ (nm) \\ \hline
        0.5 & 2.84 & 3.09 & 3.09 & 3.01 \\
        1.0 & 3.02 & 3.10 & 3.19 & 3.10 \\
        1.5 & 3.29 & 3.35 & 3.44 & 3.36 \\
        2.0 & 3.64 & 3.78 & 3.78 & 3.73
    \end{tabular}
    \caption{Discommensuration widths $d_j$ and their average $\bar{d}$ estimated from the analyses with several choices of $\sigma$ of the Gaussian used in the Lawler-Fujita method. }
    \label{tabS1}
\end{table}

In the main text, we defined the discommensurate region as the region where $n_j(\mathbf{r})$ is closer to a half-integer than to an integer.
To illustrate this region, we introduce the quantity $\delta n_j(\mathbf{r})$ given by
\begin{equation}
    \delta n_j(\mathbf{r}) \equiv |n_j(\mathbf{r}) - \mathrm{nint}[n_j(\mathbf{r})]|,
\end{equation}
which varies between 0 in the middle of the commensurate domains and 0.5 along the discommensurations.
Figures \ref{figs-width}(a)--\ref{figs-width}(c) display the images of $\delta n_j(\mathbf{r})$.
The thick white wavy strips indicate that the phase jump occurs over a finite length. 
The red curves are the contours of $n_j(\mathbf{r})$ and provide the total length of the discommensuration, $l_j$.
Figures \ref{figs-width}(d)--\ref{figs-width}(f) display the discommensurate region determined by the condition $\delta n_j(\mathbf{r}) > 0.25$, which provides the total area of the discommensurate region, $A_j$.
We estimated the average width of the discommensuration of the $j$-th CDW component as $d_j = A_j / l_j$, which is shown in Table \ref{tabS1}.
Since the $\sigma$ of the Gaussian function used in the Lawler--Fujita method serves as a broadening factor, we examined the dependence of $d_j$ on $\sigma$.
Although the obtained $d_j$ values increase with increasing $\sigma$, they are still smaller than $\sigma$, ensuring that the finite $d_j$ is not an artifact.
We believe that the intrinsic width of the discommensurate region is given by $d_j \simeq 3$ nm estimated for $\sigma = 0.5-1$ nm.

The finite width of the discommensuration does not affect the structural appearance of the domains.
The formation of the alternating triangular domains reflects the phenomenon that the discommensurations in the three directions intersect at a single vertex.
Since this phenomenon is independent of the width of the discommensurations,
the triangular structure largely remains intact even when the discommensurate regions are explicitly depicted in the domain image [Fig. \ref{figs-width}(g)].

\section{Reproducibility check on another sample}

\begin{figure*}
    \includegraphics[width=\linewidth]{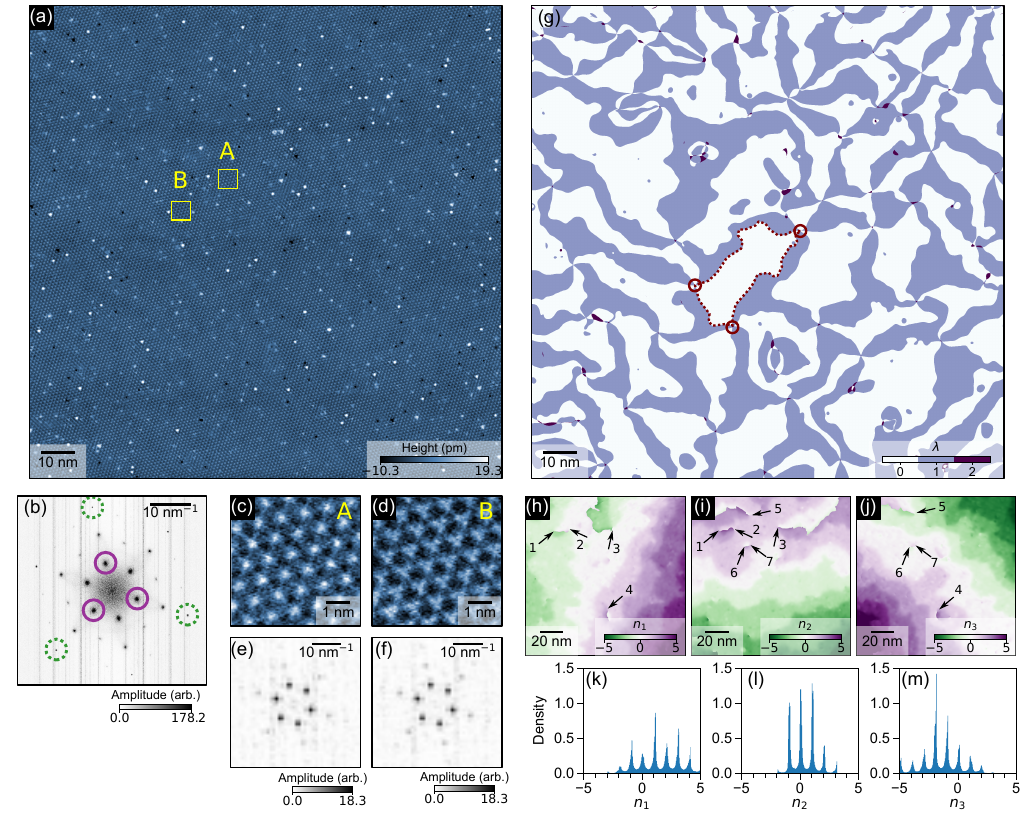}
    \caption{\label{figs-another} 
    (a) Topographic image of a 150 nm $\times$ 150 nm area of a cleaved surface of a 2\textit{H}--NbSe$_2$ crystal purchased from HQ Graphene. 
    The data were recorded at a resolution of 1024 $\times$ 1024.
    The image was flattened by subtracting a third-order polynomial fit from each line to remove the slope and a small curvature.
    The feedback condition was 100 pA at 50 mV. 
    (b) Fourier transform of the topographic image in (a).
    The dotted and solid circles indicate the spots of the atomic lattice and those of the CDW, respectively. 
    (c) and (d) Cropped images of the CC and HC domains indicated by the boxes labeled A and B in (a). 
    (e) and (f) Fourier transform images of cropped data of (c) and (d). 
    (g) Image of $\lambda({\mathbf{r}})$. 
    The CC and HC domains are depicted as white and blue areas, respectively.
    The circles and dotted curves are the vertices and edges of one of a CC domain, respectively. 
    (h)--(j) Images of $n_1(\mathbf{r})$, $n_2(\mathbf{r})$, and $n_3(\mathbf{r})$, which were obtained from the topographic data in (a) and were used to calculate the domain structure in (g). 
    The arrows indicate the locations of topological defects.
    Each pair of topological defects at the same location but in different CDW components is labeled with the same number. 
    (k)--(m) Histograms of $n_1(\mathbf{r})$, $n_2(\mathbf{r})$, and $n_3(\mathbf{r})$. 
    The vertical axis represents the frequency density.
}\end{figure*}

We have confirmed the observation of the triangular domains of CDW for STM images of different regions of the same sample, as well as for STM images of several 2\textit{H}--NbSe$_2$ samples. 
As evidence of the reproducibility, we show the analysis of the CDW domains on a 2\textit{H}--NbSe$_2$ crystal purchased from HQ Graphene [Fig. \ref{figs-another}].
The analysis was done for a topographic image of a 150 nm x 150 nm region [Fig. \ref{figs-another}(a)]. 
The Fourier transform resolves peaks of CDW and crystal lattice [Fig. \ref{figs-another}(b)]. 
The images cropped from regions labeled A and B in Fig. \ref{figs-another}(a) shows typical CC and HC structures [Figs. \ref{figs-another}(c) and \ref{figs-another}(d)] and their Fourier amplitude images are nearly identical [Figs. \ref{figs-another}(e) and \ref{figs-another}(f)].
By applying the domain visualization method proposed in our study, we obtained the domain structure shown in Fig. \ref{figs-another}(g). 
Most of the region is occupied by alternating triangular domains, supporting our main finding. 
The images of $n_1(\mathbf{r})$, $n_2(\mathbf{r})$, and $n_3(\mathbf{r})$ used to generate the domain image exhibit step-terrace structures [Figs. \ref{figs-another}(h)-\ref{figs-another}(j)].
Their histograms display sharp peaks at integer multiple of the interplanar spacing [Figs. \ref{figs-another}(k)-\ref{figs-another}(m)]. 
Figs. \ref{figs-another}(h)-\ref{figs-another}(j) also reveals the locations of topological defects as indicated by arrows. 
One can confirm that the topological defects are created in pairs in two of the three CDW components.

\section{Numerical simulation of CDW domain structure based on empirical free energy}

\subsection{Free energy for single-layer transition-metal dichalcogenides}

\begin{figure}
    \includegraphics[width=\linewidth]{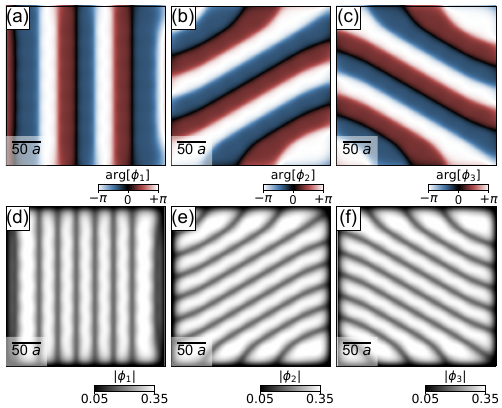}
    \caption{\label{figs-tdglop}
    Order parameters obtained by minimizing Nakanishi-Shiba's free energy.
    (a)--(c) Images of $\operatorname{arg}[\phi_1(\mathbf{r})]$, $\operatorname{arg}[\phi_2(\mathbf{r})]$, and $\operatorname{arg}[\phi_1(\mathbf{r})]$.
    (d)--(f) Images of $|\phi_1(\mathbf{r})|$, $|\phi_2(\mathbf{r})|$, and $|\phi_3(\mathbf{r})|$.
}\end{figure}

Nakanishi and Shiba proposed a free energy expression to describe the CDW transition of transition-metal dichalcogenides (TMDCs) \cite{Nakanishi1983,Nakanishi1986}.
The CDW is characterized by an incommensurate wavevector $\mathbf{Q}_j^\mathrm{IC}$ that is close to $\mathbf{Q}_j = \mathbf{b}_j/3$.
The order parameter $\phi_j(\mathbf{r})$ is a complex-valued function chosen so that the charge density attributed to the CDW is
\begin{equation}
    \rho_\mathrm{CDW}(\mathbf{r}) = \operatorname{Re}\sum_{j=1}^3 e^{i\mathbf{Q}_j \cdot \mathbf{r}} \phi_j(\mathbf{r}).
\label{eq:rhocdw}\end{equation}
We use the notation $\phi_{j+3} = \phi_j$.
The free energy $F(\phi_1, \phi_2, \phi_3)$ is given by
\begin{widetext}\begin{equation}\begin{split}
    F(\phi_1, \phi_2, \phi_3) = \int \mathrm{d}\mathbf{r} \Biggl\{
\sum_{j=1}^3 \biggl[ 
(T - T_\mathrm{CDW})|\phi_j|^2 
+ s\left| \left( -i\nabla - \mathbf{q}_j \right) \phi_j \right|^2 + B|\phi_j|^4 + C|\phi_j|^2|\phi_{j+1}|^2 + \\
+ \operatorname{Re}( Y e^{\pm iy}\phi_j^3 + W e^{\pm iw}\phi_j^2\phi_{j+1}^*\phi_{j+2}^* ) \biggr] 
+ \operatorname{Re}( D e^{\pm id}\phi_1\phi_2\phi_3 )
\Biggr\},
\end{split}\end{equation}\end{widetext}
where $\mathbf{q}_j \equiv \mathbf{Q}_j^\mathrm{IC} - \mathbf{Q}_j$, $T$ is the temperature, $T_\mathrm{CDW}$ is the CDW transition temperature, and $B$ and $C$ are real parameters.
The terms with $Y$ and $W$ are commensurability energies that regulate the phase between the CDW components, while the term with $D$ is the phase term, which adjusts the phase relation between the CDW and the crystal lattice.
The $\pm$ signs in the terms with $Y$, $W$, and $D$ correspond to the even layer ($+$) and odd layer ($-$) of the 2\textit{H} structure.
The crystal structure of the surface layer used in our study [Fig. 1(a) of the main text] corresponds to the odd layer in their definition.
The free energy of bulk TMDCs has an additional term for the interaction between layers.
The relative stability of the three types of commensurate structure depends on the parameters $y$, $w$, and $d$.
The condition where the CC and HC structures become equally stable is obtained by setting $y = w = 0$ and $d = \pi$.

Note that in the Nakanishi--Shiba papers, the origin of the spatial coordinates is considered to be a transition metal site, which differs from our convention.
In addition, their definition of the indices $n_1$, $n_2$, and  $n_3$ labeling the domains has the opposite sign to our definition of $n_1$, $n_2$, and $n_3$.
As a result, $\lambda = (n_1 + n_2 + n_3) \mod 3$ also has the opposite sign.
Consequently, the domains labeled $\lambda = 0$, 1, and 2 in their papers correspond to the TC, HC, and CC structures, respectively.

We employed a time-dependent Ginzburg-Landau approach to minimize the free energy.
The variation of $F$ with respect to $\phi_j^*$ is given by
\begin{widetext}\begin{equation}\begin{split}
    \frac{\delta F}{\delta\phi_j^*} = 
(T - T_c)\phi_j
+ s\left( -i\nabla - \mathbf{q}_j \right)^2 \phi_j 
+ 2B|\phi_j|^2\phi_j
+ C\phi_j(|\phi_{j+1}|^2 + |\phi_{j+2}|^2)
+ \frac{3}{2} Y e^{\mp iy} \phi_j^{*2} + \\
+ \frac{1}{2} W e^{\mp iw} 2\phi_j^*\phi_{j+1}\phi_{j+2} + \frac{1}{2} W e^{\pm iw}(\phi_{j+1}^2\phi_{j+2}^* + \phi_{j+2}^2\phi_{j+1}^*) 
+ \frac{1}{2} D e^{\mp id} \phi_{j+1}^*\phi_{j+2}^*.
\end{split}\end{equation}\end{widetext}
This provides the time evolution of the free energy \cite{Bray2002,Vogelgesang2017},
\begin{equation}
    \frac{\partial \phi_j}{\partial t} = -\frac{\delta F(\phi_1, \phi_2, \phi_3)}{\delta \phi_j^*}.
\end{equation}
We computed the time evolution of $\phi_j$ using this equation until the change in $F$ became sufficiently small.
We used the split-operator Fourier method to handle the terms involving spatial derivatives in the reciprocal space.
The calculations were performed on a $1024 \times 1024$ grid over a $300a \times 300a$ area ($a$ is the lattice constant).
The parameters used in the computation are $\mathbf{q}_j = -0.025\mathbf{Q}_j$, $s = 1000$, $T_\mathrm{CDW} = 1$, $T = 0.87$, $B = 2$, $C = 1$, $D = 0.2$, $Y = 0.45$, $W = 1$, $y = w = 0$, and $d = \pi$.
These parameters are the same as those used in Ref. \cite{Nakanishi1986}.

Figures \ref{figs-tdglop}(a)-\ref{figs-tdglop}(c) and \ref{figs-tdglop}(d)-\ref{figs-tdglop}(f) show the phase and amplitude of $\phi_j(\mathbf{r})$, respectively. 
In discommensurations, $\operatorname{arg}[\phi_j(\mathbf{r})]$ changes abruptly, and $|\phi_j(\mathbf{r})|$ is locally suppressed.
Since we set $\phi_j = 0$ as the boundary condition, the resulting $\phi_j(\mathbf{r})$ is distorted near the boundary. 
The topographic image is simulated as a weighted sum of $\rho_\mathrm{Se}(\mathbf{r})$ and $\rho_\mathrm{CDW}(\mathbf{r})$ given by Eqs. (\ref{eq:rhose}) and (\ref{eq:rhocdw}), respectively, with an appropriate adjustment of the origin of the spatial coordinates. 
The triplet $(n_1, n_2, n_3)$ is obtained from $\operatorname{arg}[\phi_j(\mathbf{r})]$, and the domain type is distinguished from the $\lambda$ computed from the triplet. 
Figure 4(a) in the main text is a combined plot of the simulated topographic image and the domain structure in a region cropped near the center where the distortion of $\phi_j(\mathbf{r})$ due to the boundary effect is negligible.


%